\documentclass{iopart}
\usepackage[latin9]{inputenc}
\usepackage{amssymb}
\usepackage{graphicx}
\usepackage{cite}
\begin{document}

\title[Post-selected von Neumann measurement with HG and LG pointer states]
{Post-selected von Neumann measurement with Hermite--Gaussian and Laguerre--Gaussian pointer states}

\author{Yusuf Turek$^{1,2, \star}$, Hirokazu Kobayashi$^{3, \ast}$, Tomotada Akutsu$^{4,5}$, Chang-Pu Sun$^{6, \dagger}$, and Yutaka Shikano$^{1,7,8, \circ}$}
\address{$^{1}$ Research Center of Integrative Molecular Systems (CIMoS), Institute
for Molecular Science, National Institutes of Natural Sciences, Okazaki, Aichi 444-8585, Japan}
\address{$^{2}$State Key Laboratory of Theoretical Physics, Institute of Theoretical
Physics, Chinese Academy of Sciences, and University of the Chinese
Academy of Sciences, Beijing 100190, China}
\address{$^{3}$Department of Electronic and Photonic System Engineering, Kochi University
of Technology, Tosayamada-cho, Kochi 782-0003, Japan}
\address{$^{4}$National Astronomical Observatory of Japan, Mitaka, Tokyo 181-8588,
Japan}
\address{$^{5}$Department of Astronomical Science, The Graduate University for Advanced
Studies (SOKENDAI), Mitaka, Tokyo 181-8588, Japan}
\address{$^{6}$Beijing Computational Science Research Center, Beijing 100084, China}
\address{$^{7}$Institute for Quantum Studies, Chapman University, Orange, CA 92866, USA}
\address{$^{8}$Materials and Structures Laboratory, Tokyo Institute of Technology, Yokohama 226-8503, Japan}

\ead{$^{\star}$yusufu@itp.ac.cn}
\ead{$^{\ast}$kobayashi.hirokazu@kochi-tech.ac.jp}
\ead{$^{\dagger}$cpsun@csrc.ac.cn}
\ead{$^{\circ}$yshikano@ims.ac.jp}

\date{\today}
\begin{abstract}
Through the von Neumann interaction followed by post-selection, we can
extract not only the eigenvalue of an observable of the measured system
but also the weak value. In this post-selected von Neumann measurement, 
the initial pointer state of the measuring device is assumed to be a fundamental Gaussian
wave function. By considering the optical implementation of the post-selected
von Neumann measurement, higher-order Gaussian modes can be used.
In this paper, we consider the Hermite--Gaussian (HG) and Laguerre--Gaussian
(LG) modes as pointer states and calculate the average shift of the
pointer states of the post-selected von Neumann measurement by assuming
the system observable $\hat{A}$ with $\hat{A}^{2}=\hat{I}$ and $\hat{A}^{2}=\hat{A}$
for an arbitrary interaction strength, where $\hat{I}$ represents
the identity operator. Our results show that the HG and LG pointer
states for a given coupling direction have advantages and disadvantages over the
fundamental Gaussian mode in improving the signal-to-noise ratio
(SNR). We expect that our general treatment of the weak values will be helpful
for understanding the connection between weak- and strong-measurement
regimes and may be used to propose new experimental setups with higher-order
Gaussian beams to investigate further the applications of weak measurement
in optical systems such as the optical vortex. 
\end{abstract}



\maketitle

\section{Introduction}

In a quantum measurement, observable information in the measured
system can be extracted from the statistical average shift of a
pointer. In this process, von Neumann interaction is initially used with the standard
model of quantum measurement by mathematically describing the coupling
between the measured system and measuring devices \cite{Neumann}.
However, such strong measurements are not time symmetric. When
considering time-symmetric quantum measurements, post-selection
of the measured system is required after the measurement interaction \cite{Aharonov(1964)}.
On summing the post-selections, the statistical average shift of the
pointer can be determined in the standard model of quantum measurement.
Therefore, throughout the present work, measurements with post-selection
are called post-selected von Neumann quantum measurements.
A particular case of post-selected von Neumann quantum measurements with sufficiently weak coupling between the measuring device and measured
system is called the weak measurement, as proposed by Aharonov, Albert,
and Vaidman (AAV) \cite{Aharonov(1988)}. This statistical average
shift of the pointer is characterized by the weak value of the observable
in the measured system \cite{Nori}.

A significant feature of the weak measurements is that the weak value
of the measured quantity can lie outside the usual range of eigenvalues
of an observable applicable for
a standard quantum measurement \cite{Aharonov(1988)}. This feature
is usually referred to as the amplification effect for weak signals
and is different from conventional quantum measurement, in which a
coherent superposition of quantum states is collapsed \cite{Neumann}. A large
weak value can amplify small unknown parameters for detecting various
properties such as beam deflection \cite{Pfeifer(2011),Hosten2008,Hogan(2011),Zhou(2013),Starling(2009),Dixon(2009)},
frequency shifts \cite{Starling(2010)-1}, phase shifts \cite{Starling(2010)},
angular shifts \cite{Magana(2013),Bertulio(2014)}, velocity shifts
\cite{viza(2013)}, and even temperature shifts \cite{Egan(2012)}.
However, the advantages of the weak-value amplification are purely technical
\cite{Feizpour(2011),Jordan(2014),Tanaka(2013),Knee(2014),Lee(2014),Knee(2014)-2,Combes(2014),Ferrie(2014),Matsuoka(2015)}.
This is based on the single parameter estimation theory.
In general, the weak value is a complex number. Thus, weak measurements
are ideal for examining the fundamentals of quantum
physics such as quantum paradoxes (Hardy's paradox \cite{Aharonov(2002),Lundeen(2009),Yokota(2009),Shikano(2010)}
and the three-box paradox \cite{Resch(2004)}), quantum correlation
and quantum dynamics \cite{Aharonov(2005),Dressel(2014),Aharonov(2008),Holger(2013),Shikano(2012),Aharonov(2011),Shikano(2011),Shikano(2011)-2,Kagami(2011)},
and quantum-state tomography \cite{Lundeen(2011),Lundeen(2012),Braveman(2013),Kocsis(2011),Malik(2014),Salvail(2013)},
as well as the violation of the generalized Leggett--Garg inequalities
\cite{Palacios,Suzuki(2012),Dressel(2011),Emary(2014),Goggin(2011),Groen(2013)}
and the violation of the initial Heisenberg measurement--disturbance
relationship \cite{Lee(2012),Eda(2014)}.

Thus far, most studies on weak measurement use the zero-mean Gaussian
state as an initial pointer state and expand the unitary
operator of evolution up to the first order because, in the weak measurement scheme,
the coupling between the measured system and measuring device is very
weak. However, when considering the connection between weak and strong
measurements, amplification limit, and measurement back-action
of the weak measurement scheme, the full-order effects of unitary
evolution due to the von Neumann interaction between the measured
system and measuring device are required. The measurements of arbitrary
coupling strength beyond the first-order interaction have been previously
discussed by Aharonov and Botero \cite{Aharonov(2005)-1}. Di Lorenzo
and Egues \cite{Lorenzo(2008)} investigated von Neumann-type
measurement to clarify detector dynamics in the weak-measurement
process. Wu and Li \cite{Wu(2011)} proposed a general formulation
of weak measurement that includes second-order effects of the
unitary evolution due to the von Neumann interaction between the system
and detector, and they theoretically demonstrated on the basis of
the second-order calculation that the back-action
effect is important in the weak-value amplification. Recently, several studies \cite{Koike(2011),Zhu(2011),Nakamura(2012)}
analytically showed that an upper bound of the weak-value amplification
exists in the post-selected von Neumann measurement by assuming that
the probe-state wave function is Gaussian and that the observable $\hat{A}$
satisfies $\hat{A}^{2}=\hat{I}$, where $\hat{I}$ is the identity
operator. On the other hand, there is no upper bound on the weak-value amplification 
on the optimal probe-state wave function~\cite{Susa(2012),DiLorenzo(2013),Susa(2013)} 
while it is so difficult to implement this wave function~\cite{Shikano2(2014)}.

In optical experiments, we encounter higher-order Gaussian
beams such as Hermite--Gaussian (HG) and Laguerre--Gaussian (LG) beams,
which are higher-order solutions of the paraxial wave equation with
rectangular and cylindrical symmetry about their axes of propagation,
respectively. Both HG and LG beams are widely used in the theory of
lasers and resonators \cite{Kogelnik(1966),Siegman(1986)}. In fact,
the zero-mean Gaussian beam is a special case of HG and LG beams.
The weak measurement with the higher-order Gaussian-beam pointer state
has been discussed in Refs. \cite{Shikano(2012)-1,Puentes(2012),Dressel2012,Shikano(2014),deLima(2014)}.
In particular, de Lima Bernardo et al. \cite{deLima(2014)} presented
a simplified algebraic description of the weak measurements with
HG and LG pointer states. In Ref. \cite{deLima(2014)},
the unitary evolution operator is considered only up to the first order, raising
an intriguing question as to whether the higher-order Gaussian beams are
more advantageous in quantum measurement compared to the fundamental Gaussian
beam.

In the present study, we determine the post-selected von Neumann quantum measurement
for an arbitrary coupling strength with HG- and LG-mode pointer states
under the assumption that the system observable $\hat{A}$ satisfies
$\hat{A}^{2}=\hat{I}$ and $\hat{A}^{2}=\hat{A}$ (projection operator).
To clarify the practical advantages of higher-order Gaussian beams,
we investigate the signal-to-noise ratio (SNR) while considering
the post-selection probability, which is defined by 
\begin{equation}
SNR_{W}=\frac{\sqrt{NP_{s}}\vert\langle W\rangle_{fi}\vert}{\sqrt{\langle W^{2}\rangle_{f}-\langle W\rangle_{f}^{2}}},\ \ \hat{W}=\hat{X},\hat{Y}.\label{eq:SNR}
\end{equation}
Here, $\langle.\rangle_{f}$ denotes the expectation value of the
measuring system operator under the final state of the pointer, and
$\hat{X}=\int x \left|x\right\rangle \left\langle x\right| dx$ ($x$ is the
coupling direction of the von Neumann measurement) and $\hat{Y}=\int y \left|y\right\rangle \left\langle y\right| dy$
($y$ is the orthogonal coupling direction). Here, $P_{s}$ is the probability that 
the post-selected state is included in the pre-selection state, and $N$ 
is the number of measurement time.
To verify our general formulas, two special limits are considered. If
the zero-mean Gaussian pointer is used as the initial state,
our general expectation values are found to reduce to the results given in
Refs. \cite{Nakamura(2012),Wu(2011)}. On the other hand, if
the evaluation is considered only up to the first order, our general expectation
values reproduce all results given in Ref. \cite{deLima(2014)}.

The remainder of this paper is organized as follows. In Section~\ref{sec2},
we present the model setup for the post-selected von Neumann measurement.
In Sections \ref{sec3} and \ref{sec4}, we first present the expressions
of HG- and LG-mode pointer states in the Fock-state representation in accordance
with de Lima Bernardo et al. \cite{deLima(2014)}. We then present general
forms of the expectation values and discuss the SNRs with HG- and LG-mode
pointer states for the system operator $\hat{A}$ with $\hat{A}^{2}=\hat{I}$
and $\hat{A}^{2}=\hat{A}$, which were used in several optical implementation on the 
weak measurement~\cite{Pfeifer(2011),Hosten2008,Hogan(2011),Zhou(2013),Starling(2009),Dixon(2009),Starling(2010)-1,Starling(2010),Magana(2013),Bertulio(2014),viza(2013),Egan(2012),Resch(2004),Suzuki(2012),Lee(2012),Eda(2014),Shikano(2014)}. 
In section \ref{sec5}, to check the
validity of our general results, we consider some special initial pointer
states and approximated treatments used in previous works and
show that our general formulas can reproduce all the related results
reported in those previous works \cite{deLima(2014),Wu(2011),Nakamura(2012)}.
We present the conclusions and remarks of our study in the final section \ref{sec6}.
Throughout this paper, we use $\hbar=1$ units.

\section{Model setup}
\label{sec2} For the post-selected von Neumann measurement, the coupling
interaction between the system and detector is considered with the standard
von Neumann Hamiltonian: 
\begin{eqnarray}
H & = & g\delta(t-t_{0})\hat{A}\otimes\hat{P_{x}},\label{eq:Hamil}
\end{eqnarray}
where $g$ is a coupling constant and $\hat{P}_{x}$ is the conjugate
momentum operator for the position operator $\hat{X}$ of the measurement
device; i.e., $[\hat{X},\hat{P_{x}}]=i\hat{I}$. We have taken the
interaction to be impulsive at time $t=t_{0}$ for simplicity. The time-evolution operator for
such impulsive interaction is $e^{-ig\hat{A}\otimes\hat{P_{x}}}$.

The post-selected von Neumann measurement is characterized by the pre-
and post-selection of the system state. If we prepare an initial
state $\left|\psi_{i}\right\rangle $ of the system and pointer
state, after some interaction time $t_{0}$, we post-select a system
state $\left|\psi_{f}\right\rangle $ and obtain information on
a physical quantity $\hat{A}$ from the pointer wave function by using the
following weak value: 
\begin{equation}
\langle A\rangle_{w}=\frac{\left\langle \psi_{f}\right|\hat{A}\left|\psi_{i}\right\rangle }{\left\langle \psi_{f}\right|\psi_{i}\rangle}.\label{eq:WV}
\end{equation}
In general, the weak value is a complex number. It is evident from Eq. $\left(\ref{eq:WV}\right)$,
that when the pre-selected state $\left|\psi_{i}\right\rangle $
and the post-selected state $\left|\psi_{f}\right\rangle $ are nearly
orthogonal to each other, the absolute value of the weak value can be arbitrarily
large, resulting in the weak-value amplification.

From the above definitions, we note that the unitary evolution operator
$e^{-ig\hat{A}\otimes\hat{P}_{x}}$ for the operator $\hat{A}$ satisfies
the property $\hat{A}^{2}=\hat{I}$ as follows: 
\begin{equation}
e^{-ig\hat{A}\otimes\hat{P}_{x}}=\frac{1}{2}\left(\hat{I}+\hat{A}\right)\otimes D\left(\frac{s}{2}\right)+\frac{1}{2}\left(\hat{I}-\hat{A}\right)\otimes D\left(-\frac{s}{2}\right).\label{eq:UNA1}
\end{equation}
Similarly, for the property $\hat{A}^{2}=A$, the evolution operator
satisfies 
\begin{equation}
e^{-ig\hat{A}\otimes\hat{P}_{x}}=\left(\hat{I}-\hat{A}\right)\otimes\hat{I}+\hat{A}\otimes D\left(\frac{s}{2}\right).\label{eq:UNA2}
\end{equation}
Here, we use the position operators $\hat{X}$ and $\hat{Y}$ as well
as their corresponding momentum operators $\hat{P}_{x}$ and $\hat{P}_{y}$,
which can be written in terms of the annihilation (creation) operators
$\hat{a}_{i}$($\hat{a}_{i}^{\dagger}$) with $i=x,y$ as \cite{Tannoudji(2005)} 
\begin{eqnarray}
\hat{X} & = & \sigma\left(\hat{a}_{x}^{\dagger}+\hat{a}_{x}\right),\label{eq:annix}\\
\hat{Y} & = & \sigma\left(\hat{a}_{y}^{\dagger}+\hat{a}_{y}\right),\label{eq:creax}\\
\hat{P}_{x} & = & \frac{i}{2\sigma}\left(\hat{a}_{x}^{\dagger}-\hat{a}_{x}\right),\label{eq:anniy}\\
\hat{P}_{y} & = & \frac{i}{2\sigma}\left(\hat{a}_{y}^{\dagger}-\hat{a}_{y}\right).\label{eq:creay}
\end{eqnarray}
Here, $\sigma$ is the width of the fundamental
Gaussian beam. It is worth noting that in these definitions, the
propagation direction of the beam is assumed to be fixed ~\cite{Allen(1993)}.
These annihilation (creation) operators satisfy the commutation relations
$\left[\hat{a}_{i},\hat{a}_{j}^{\dagger}\right]=\delta_{ij}\hat{I}$
with $i,j=x,y$. The parameter $s$ is defined as $s:\equiv g/\sigma$,
and $D\left(\xi\right)$ is a displacement operator with complex $\xi$
defined as
\begin{equation}
D(\xi)=e^{\xi\hat{a}_{x}^{\dagger}-\xi^{\ast}\hat{a}_{x}}, \label{eq:DOP}
\end{equation}
Here, the parameter $s$ characterizes the measurement strength. Note that the interaction 
between the system and pointer is weak (strong) if $s\ll1$ $\left(s\gg1\right)$.

In the following sections, we consider the post-selected von Neumann measurement
with HG- and LG-mode pointer states for an arbitrary measurement-strength
parameter $s$ for the system operator $\hat{A}$ with $\hat{A}^{2}=\hat{I}$
and $\hat{A}^{2}=\hat{A}$, respectively. On the choice of the system operator $\hat{A}$, 
$\hat{A}^{2}=\hat{I}$ and $\hat{A}^{2}=\hat{A}$ are taken as the qubit operator and 
the projector, respectively.

\section{Post-selected von Neumann measurements with HG-mode pointer states}
\label{sec3} The general HG modes can be generated from the fundamental
Gaussian mode, $\left|0,0\right\rangle _{HG}$, and can be defined
as\cite{Tannoudji(2005),deLima(2014)} 
\begin{equation}
\left|n,m\right\rangle _{HG}=\frac{1}{\sqrt{n!m!}}\left(\hat{a}_{x}^{\dagger}\right)^{n}\left(\hat{a}_{y}^{\dagger}\right)^{m}\left|0,0\right\rangle _{HG}.\label{eq:HG mode}
\end{equation}
These modes are complete sets of solutions to the paraxial wave equation
in rectangular coordinates. Any arbitrary paraxial wave can be described
as a superposition of HG modes with the appropriate weighting and
the phase factors. Practically, the higher-order HG modes can be simply
generated by inserting cross wires into the laser cavity with the
wires aligned with the nodal lines of the desired HG mode \cite{Am(1996),OPt(1993)}.
However, a more convenient way for generating higher-order modes is
the use of computer-generated holograms or a spatial light modulator
(SLM) \cite{LaserP(2011)}, which allows reprogrammable waveform generation
controlled using a computer.

In the present paper, the initial state of the HG-mode pointer is considered to be
$\left|\phi_{i}\right\rangle =\left|n,m\right\rangle _{HG}$. Note
that the HG modes can be factored in functions that depend on $x$
and $y$ directions. In our standard von Neumann measurement Hamiltonian
$\left(\ref{eq:Hamil}\right)$, only $x$-direction interaction exists;
thus, the $y$-direction quantum number $m$ is omitted in the HG-mode calculations.

In what follows, we discuss the post-selected von Neumann measurement
for the system operator $\hat{A}$ that satisfies the properties $\hat{A}^{2}=\hat{I}$
and $\hat{A}^{2}=\hat{A}$.

\subsection{$\hat{A}^{2}=\hat{I}$ case}
After the unitary evolution given in Eq. $\left(\ref{eq:UNA1}\right)$,
the system state is post-selected to $\left|\psi_{f}\right\rangle $.
Then, we obtain the following normalized final pointer states: 
\begin{equation}
\left|\phi_{f_{1}}\right\rangle =\frac{\lambda}{2}\left[D\left(-\frac{s}{2}\right)+D\left(\frac{s}{2}\right)+\langle A\rangle_{w}\left(D\left(\frac{s}{2}\right)-D\left(-\frac{s}{2}\right)\right)\right]\left|n\right\rangle _{HG},\label{eq:HGA2-1}
\end{equation}
where the normalization coefficient is given by 
\begin{equation}
\lambda=\left[1+\frac{1}{2}\left(1-\vert\langle A\rangle_{w}\vert^{2}\right)\left(e^{-\frac{s^{2}}{2}}L_{n}\left(s^{2}\right)-1\right)\right]^{-\frac{1}{2}}.\label{eq:HGA21co}
\end{equation}
Here, the Laguerre polynomials are defined as 
\begin{equation}
L_{n}\left(x\right)=\sum_{\epsilon=0}^{n}\left(\begin{array}{c}
n\\
\epsilon
\end{array}\right)\frac{(-1)^{\epsilon}}{\epsilon!}x^{\epsilon}.
\end{equation}
The explicit expression of Eq. (\ref{eq:HGA2-1}) can be obtained
using the displaced Fock states defined as \cite{Knight(1990),Roy(1982)}
\begin{equation}
D(\xi)\left|n\right\rangle _{HG}=e^{-\frac{\vert\xi\vert^{2}}{2}}\sum_{\kappa=0}^{\infty}\left(\frac{n!}{\kappa!}\right)^{\frac{1}{2}}\left(\xi\right)^{\kappa-n}L_{n}^{(\kappa-n)}\left(\vert\xi\vert^{2}\right)\left| \kappa \right\rangle .\label{eq:DFS}
\end{equation}
Here, the generalized Laguerre polynomials are defined as 
\begin{equation}
L_{n}^{(\eta)}\left(x\right)=\sum_{i=0}^{n}\left(\begin{array}{c}
n+\eta\\
n-i
\end{array}\right)\frac{(-1)^{i}}{i!}x^{i},
\end{equation}
where $\eta$ is an integer. Using Eqs. $\left(\ref{eq:HGA2-1}\right)$ and
$\left(\ref{eq:DFS}\right)$, we can calculate the general forms of
the expectation values of the conjugate momentum $\hat{P}_{x}$ and
position operator $\hat{X}$ under the final pointer states $\left|\phi_{f_{1}}\right\rangle $,
which are given by 
\begin{equation}
\langle X\rangle_{f_{1}}^{HG}=\vert\lambda\vert^{2}g\Re\langle A\rangle_{w}\label{eq:HGGXA21}
\end{equation}
and 
\begin{eqnarray}
2g\langle P_{x}\rangle_{f_{1}}^{HG} & = & \vert\lambda\vert^{2}s^{2}\Im\langle A\rangle_{w}e^{-\frac{s^{2}}{4}}\times\label{eq:HGGPA21}\\
 &  & \sum_{\kappa=0}^{\infty}\frac{n!(\frac{-s^{2}}{4})^{\kappa-n}}{\kappa!}L_{n}^{(\kappa-n)}\left(\frac{s^{2}}{4}\right)L_{n}^{(\kappa-n+1)}\left(\frac{s^{2}}{4}\right),\nonumber 
\end{eqnarray}
respectively. Eqs. $\left(\ref{eq:HGGXA21},\ref{eq:HGGPA21}\right)$
are the general forms of expectation values for the system operator
$\hat{A}$ satisfying $\hat{A}^{2}=\hat{I}$, and they are valid for
an arbitrary value of the measurement-strength parameter $s$.

To investigate the practical advantages of the higher-order Gaussian modes,
we check the signal-to-noise ratio (SNR) in
two cases. Here, we consider the two-dimensional quantum (qubit) state
and assume that the operator $\hat{A}$ to be observed is the
$x$-component of the spin of a spin-$1/2$ particle through the von Neumann interaction
$\left(\ref{eq:Hamil}\right)$ 
\begin{equation}
\hat{A}=\hat{\sigma}_{x}=\left|\uparrow_{z}\right\rangle \left\langle \downarrow_{z}\right|+\left|\downarrow_{z}\right\rangle \left\langle \uparrow_{z}\right|.\label{eq:A2 sigma}
\end{equation}
Here, $\left|\uparrow_{z}\right\rangle $ and $\left|\downarrow_{z}\right\rangle $
are eigenstates of $\hat{\sigma}_{z}$ with corresponding eigenvalues of
$1$ and $-1$, respectively. We select the pre- and post-selected
states as 
\begin{equation}
\left|\psi_{i}\right\rangle =\cos\left(\frac{\theta}{2}\right)\left|\uparrow_{z}\right\rangle +e^{i\phi}\sin\left(\frac{\theta}{2}\right)\left|\downarrow_{z}\right\rangle \label{eq:pres}
\end{equation}
and 
\begin{equation}
\left|\psi_{f}\right\rangle =\left|\uparrow_{z}\right\rangle ,\label{eq:posts}
\end{equation}
respectively. Thus, we can obtain the weak value by substituting these
states into Eq. $\left(\ref{eq:WV}\right)$: 
\begin{equation}
\langle A\rangle_{w}=e^{i\phi}\tan\frac{\theta}{2},\label{eq:wv}
\end{equation}
where $\theta\in[0,\pi]$ and $\phi\in[0,2\pi).$ Here, the probability of 
post-selection is $P_{s}=\cos^{2}\left(\theta/2\right)$. Throughout 
the present paper, these pre- and post-selected states are used in the 
analysis of SNRs. 

\begin{figure}[ht]
\begin{center}
\includegraphics[width=4.5cm]{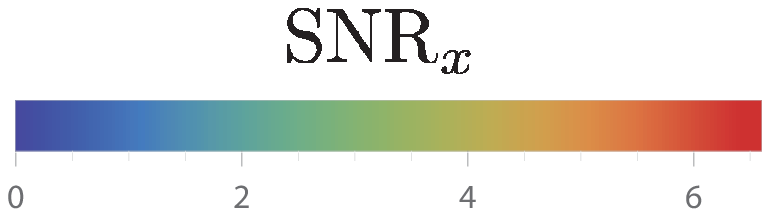}

\includegraphics[width=4cm]{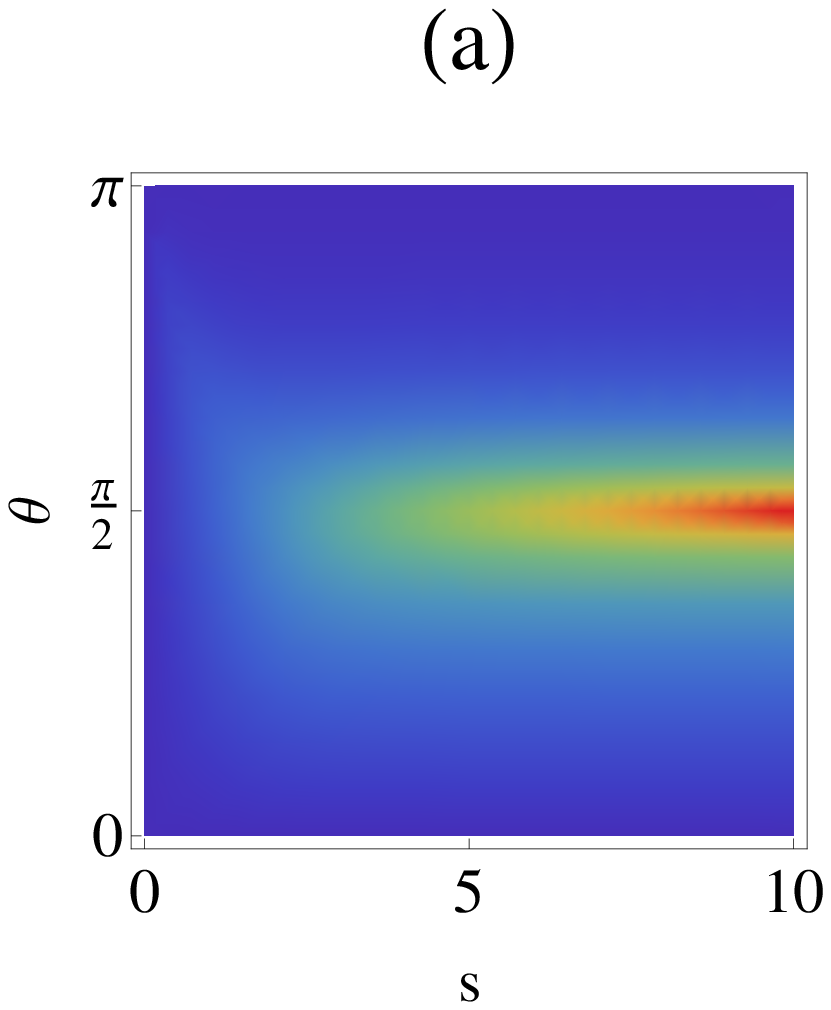}\includegraphics[width=4cm]{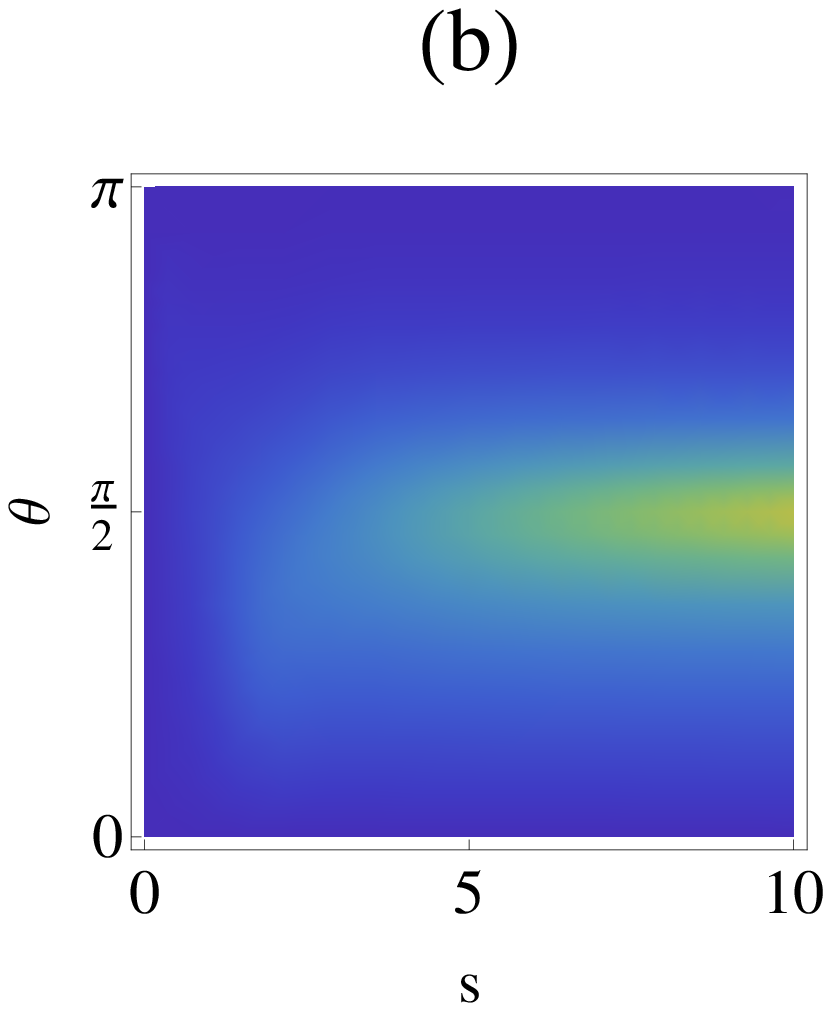}\includegraphics[width=4cm]{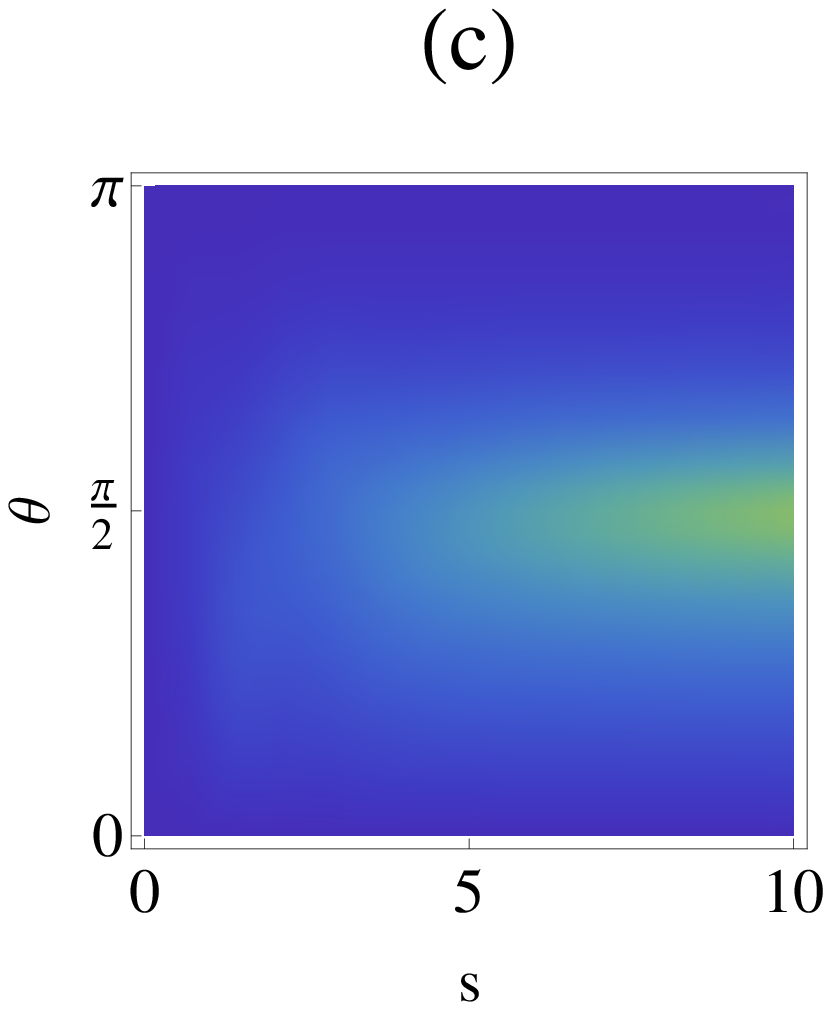}
\end{center}
\protect\caption{(Color online)\label{fig:SNRA21HG} SNR in the $x$-direction
for HG-mode pointer states with the operator $\hat{A}$ satisfying
the property $\hat{A}^{2}=\hat{I}$ plotted with respect to the
measurement-strength parameter $s$ and pre-selection angle $\theta$
for the mode (a) $n=0$, (b) $n=1$, and (c) $n=2$.
We use $\phi=0$ in Eq. $\left(\ref{eq:wv}\right)$ in all figures.}
\end{figure}

In Fig. $\ref{fig:SNRA21HG}$, the behaviour of the SNR is shown as
a function of the measurement-strength parameter $s$ and pre-selection
angle $\theta$. When $\phi=0$, the weak value becomes $\tan\frac{\theta}{2}$.
We can see that the SNR decreases as $n$ increases (higher-order
modes). A ridge exists around $\theta=\pi/2$, which is a result
of strong measurement; when $\theta=\pi/2$, the pre-selection state
is the eigenstate of the operator $\hat{\sigma}_{x}$ with the corresponding
eigenvalue $+1$. In Fig. $\ref{fig:SNRA21HG}$, we can also identify
a bridge between the weak measurement regime ($s\ll1$) and strong
measurement regime ($s\gg1$). As the SNR is proportional to the root of 
the measurement time, we consider $N=1$ throughout this paper. 
These results show that the fundamental Gaussian pointer state is better 
than the other HG modes on the improvement of the SNR. 

\begin{figure}[t]
\begin{center}
\includegraphics[width=6cm]{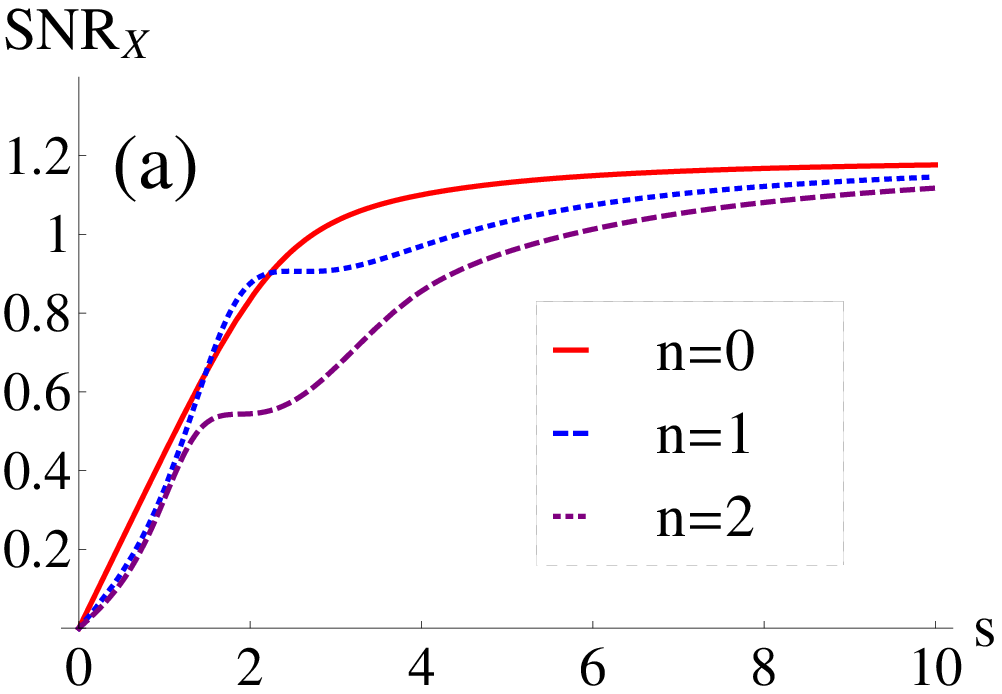} \ \includegraphics[width=6cm]{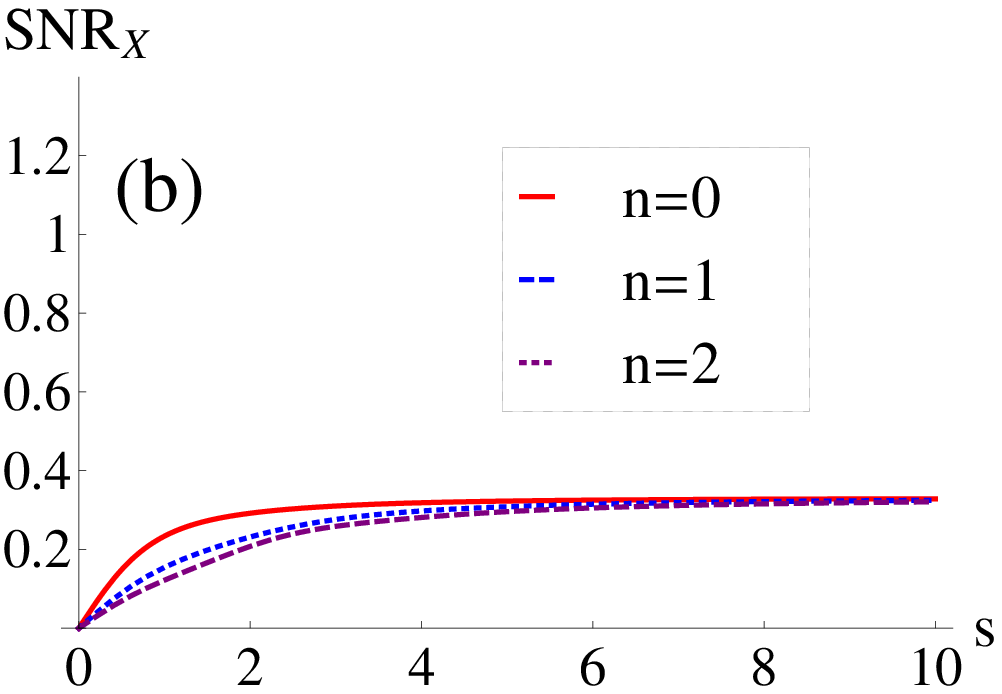}

\includegraphics[width=6cm]{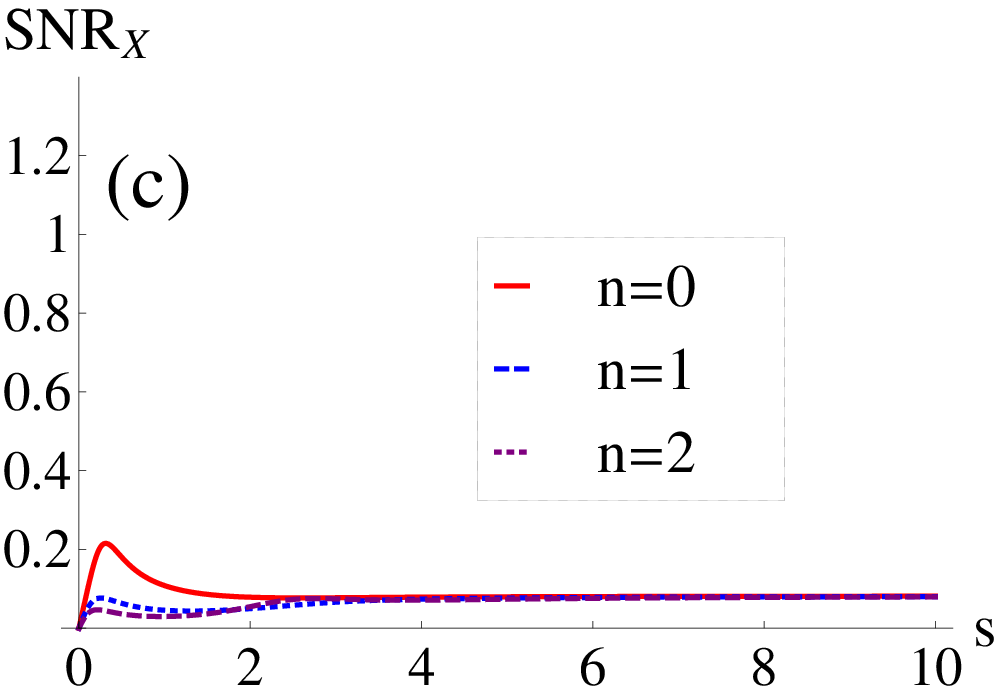} \ \includegraphics[width=6cm]{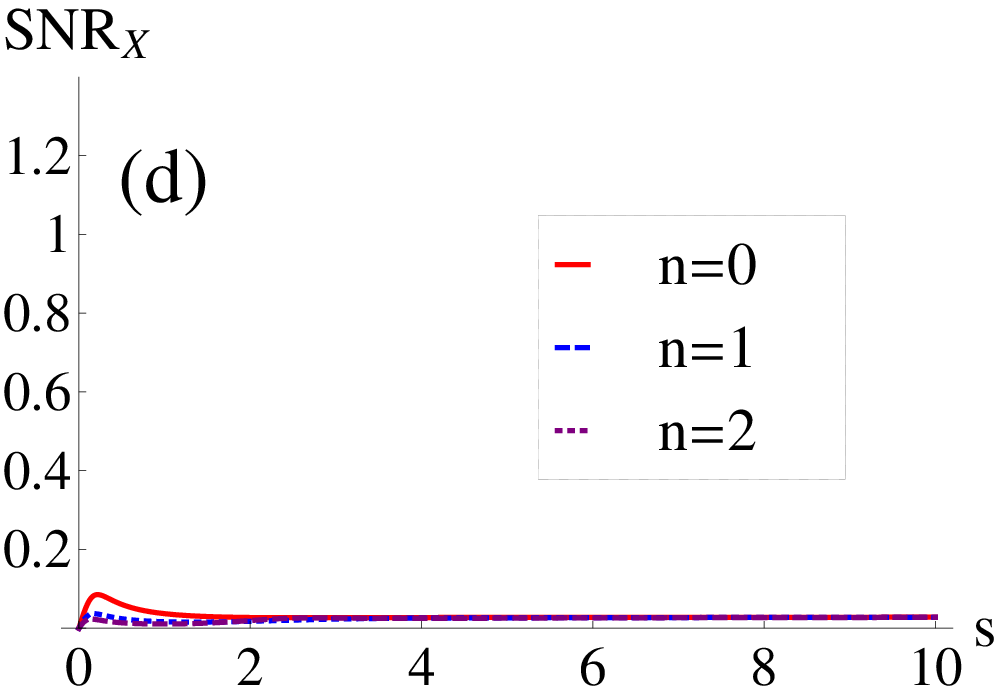}
\end{center}
\protect\caption{(Color online)\label{fig:SNRA21HG2D} SNR in the $x$-direction
for HG-mode pointer states with the operator $\hat{A}$ satisfying
the property $\hat{A}^{2}=\hat{I}$ plotted with respect to the
measurement-strength parameter $s$ for some specific weak values:
(a) $\langle A\rangle_{w}=0.5$, (b) $\langle A\rangle_{w}=0.5+i$,
(c) $\langle A\rangle_{w}=5$, and (d) $\langle A\rangle_{w}=5+5i$~\cite{paraex}.}
\end{figure}

We also check the SNR with some specific weak values, and the numerical
results are given in Fig. $\ref{fig:SNRA21HG2D}$. As shown in Figs.
$\ref{fig:SNRA21HG}$ and $\ref{fig:SNRA21HG2D}$, the higher-order
HG modes have no practical advantages in improving the SNR. We also
note that the imaginary part of the weak value has no role in improving
the SNR in the $x$-direction. These results are in general supported by 
Refs.~\cite{Knee(2014)-2,Justin}.

\subsection{$\hat{A}^{2}=\hat{A}$ case}
\begin{figure}
\begin{center}
\includegraphics[width=6cm]{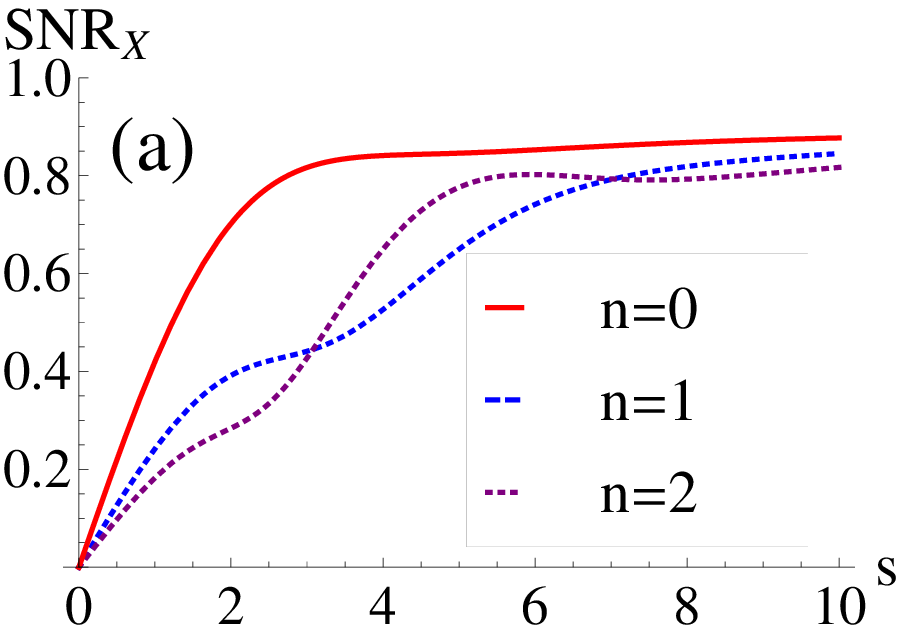} \ \includegraphics[width=6cm]{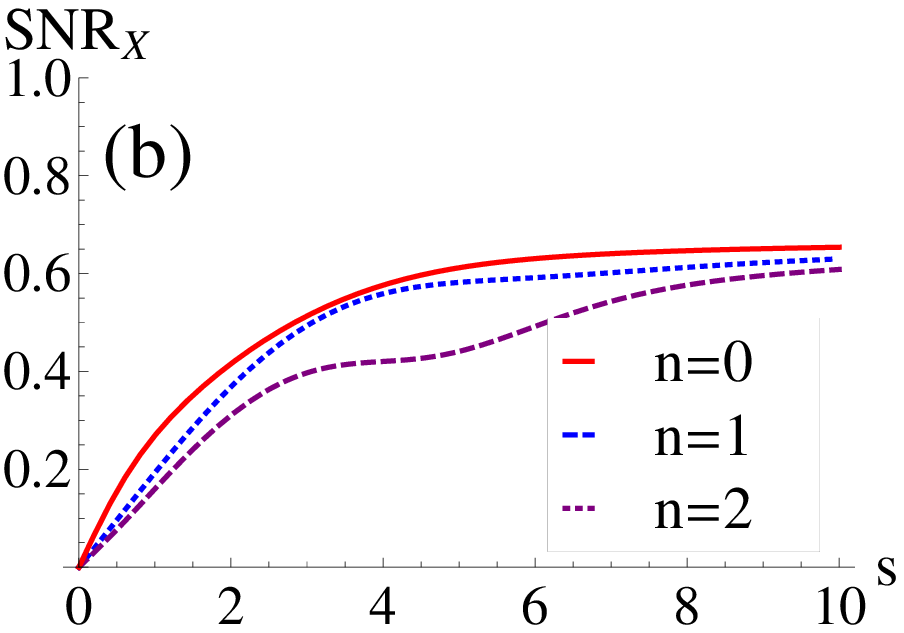}

\includegraphics[width=6cm]{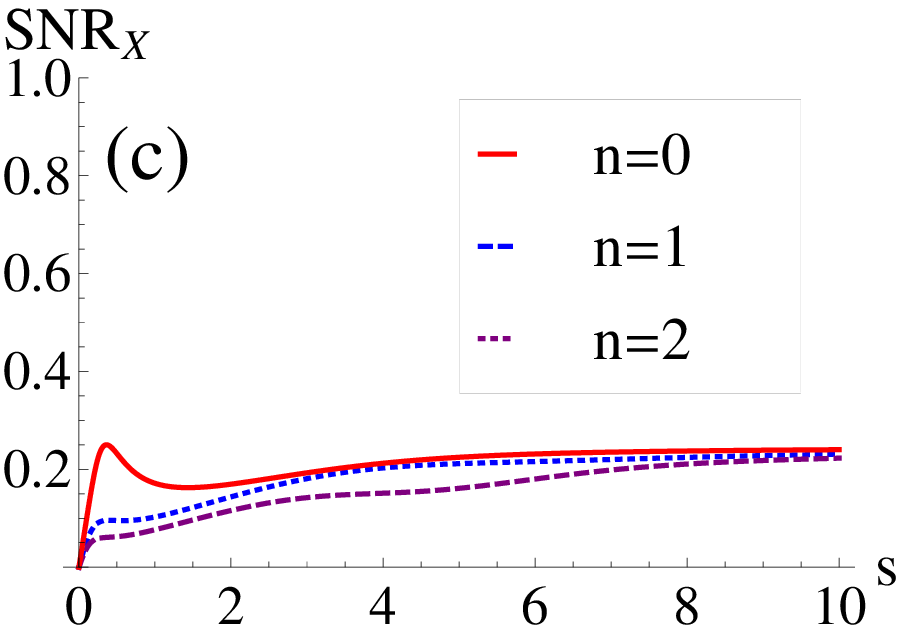} \ \includegraphics[width=6cm]{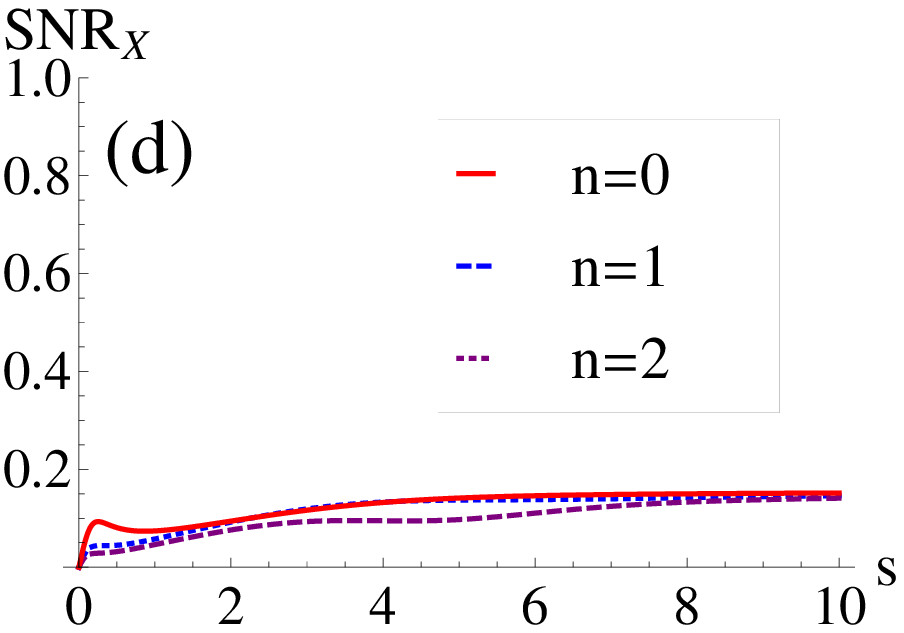}
\end{center}
\protect\caption{(Color online)\label{fig:SNR A2AHG} SNR in the $x$-direction
for HG-mode pointer states with the operator $\hat{A}$ satisfying
the property $\hat{A}^{2}=\hat{A}$ plotted with respect to the
measurement-strength parameter $s$ for some specific weak values:
(a) $\langle A\rangle_{w}=0.5$, (b) $\langle A\rangle_{w}=0.5+i$,
(c) $\langle A\rangle_{w}=5$, and (d) $\langle A\rangle_{w}=5+5i$~\cite{paraex}.}
\end{figure}

By following the process used for the $\hat{A}^{2}=\hat{I}$ case, we can
obtain the normalized final pointer states after the unitary evolution
given in Eq. $\left(\ref{eq:UNA2}\right)$. The post-selection to
$\left|\psi_{f}\right\rangle $ is given as follows: 
\begin{equation}
\left|\phi_{f_{2}}\right\rangle =\gamma\left[1-\langle A\rangle_{w}+\langle A\rangle_{w}D\left(\frac{s}{2}\right)\right]\left|\phi_{i}\right\rangle ,\label{eq:HGA2A}
\end{equation}
where $\gamma$ is the normalization coefficient given by 
\begin{equation}
\gamma=\left[1+2\left(\Re\langle A\rangle_{w}-\vert\langle A\rangle_{w}\vert^{2}\right)\left(e^{-\frac{s{}^{2}}{8}}L_{n}\left(\frac{s^{2}}{4}\right)-1\right)\right]^{-\frac{1}{2}}.\label{eq:HGA2Aco}
\end{equation}
Thus, by using Eqs. $(\ref{eq:DFS})$ and $(\ref{eq:HGA2A})$, we can
calculate the general forms of expectation values of the conjugate
momentum $\hat{P}_{x}$ and position operator $\hat{X}$ under
the final pointer sates $\left|\phi_{f_{2}}\right\rangle $; the obtained
results are given by 
\begin{equation}
\langle X\rangle_{f_{2}}^{HG}=\vert\gamma\vert^{2}g\left(\Re\langle A\rangle_{w}-\vert\langle A\rangle_{w}\vert^{2}\right)e^{-\frac{S^{2}}{8}}L_{n}\left(\frac{s^{2}}{4}\right)+\vert\gamma\vert^{2}g\vert\langle A\rangle_{w}\vert^{2}\label{eq:HGA2AX}
\end{equation}
and 
\begin{equation}
2g\langle P_{x}\rangle_{f_{2}}^{HG}=\vert\gamma\vert^{2}s^{2}\Im\langle A\rangle_{w}e^{-\frac{s^{2}}{8}}\left(L_{n}^{(1)}\left(\frac{s^{2}}{4}\right)+L_{n-1}^{(1)}\left(\frac{s{}^{2}}{4}\right)\right),\label{eq:HGA2AP}
\end{equation}
respectively. In these calculations, we use the following properties
of the displaced Fock states \cite{Ferrano(2005)}:
\begin{eqnarray}
 & _{HG}\left\langle n\!\!+d\right|D\left(\xi\right)\left|n\right\rangle _{HG} =\sqrt{\frac{n!}{(n+d)!}}e^{-\frac{\vert\xi\vert^{2}}{2}}\xi^{d}L_{n}^{(d)}\left(\vert\xi\vert^{2}\right),\\
 & _{HG}\left\langle n\right|D\left(\xi\right)\left|n\!\!+\!\! d\right\rangle _{HG} =\sqrt{\frac{n!}{\left(n+d\right)!}}e^{-\frac{\vert\xi\vert^{2}}{2}}\!\!(-\xi^{\ast})^{d}L_{n}^{(d)}\left(\vert\xi\vert^{2}\right),\\
 & _{HG}\left\langle n\right|D\left(\xi\right)\left|n\right\rangle _{HG}  =e^{-\frac{\vert\xi\vert^{2}}{2}}L_{n}\left(\vert\xi\vert^{2}\right).
\end{eqnarray}
We know that the operator $\hat{A}$ satisfying
the property $\hat{A}^{2}=\hat{A}$ can be a projection operator $\hat{A}=\left|C\right\rangle \left\langle C\right|$
that can also be taken as $\hat{A}=(\hat{I}\pm\hat{B)}/2$ with $\hat{B}^{2}=\hat{I}$.
This type of operator has numerous applications in the weak measurement
theory, such as in the three-box paradox problem \cite{Resch(2004)}
and quantum tomography~\cite{Lundeen(2011),Lundeen(2012)}. In the present
paper, we consider $\hat{A}=(\hat{I}+\hat{\sigma_{x}})/2$ and choose
the pre- and post-selected states as given in Eq. $\left(\ref{eq:pres}\right)$ and
Eq. $\left(\ref{eq:posts}\right)$, respectively. The numerical results
are shown in Fig. $\ref{fig:SNR A2AHG}$. As indicated in Fig. $\ref{fig:SNR A2AHG}$,
the higher-order HG modes have no practical advantages in improving
the SNR for the operator $\hat{A}$ satisfying the property $\hat{A}^{2}=\hat{A}$,
and the imaginary part of the weak values has no role in increasing
the SNR, as mentioned above. 

\section{Post-selected von Neumann measurements with LG-mode pointer states}
\label{sec4} The general LG modes can be defined as~\cite{deLima(2014),Tannoudji(2005)}
\begin{equation}
\left|\mu,\nu\right\rangle _{LG}=\left(\frac{1}{2}\right)^{\frac{\alpha+\beta}{2}}\frac{1}{\sqrt{\alpha!\beta!}}\left(\hat{a}_{x}^{\dagger}+i\hat{a}_{y}^{\dagger}\right)^{\alpha}\left(\hat{a}_{x}^{\dagger}-i\hat{a}_{y}^{\dagger}\right)^{\beta}\left|0,0\right\rangle _{HG},\label{eq:LG mode}
\end{equation}
where $\nu$ and $\mu$ are integers. Here, the indices $\alpha=(\mu+\nu)/2$
and $\beta=(\mu-\nu)/2$ are related to the usual radial and
azimuthal indices ($p$ and $l$, respectively) by the relations $p=\min\left(\alpha,\beta\right)$
and $l=\vert\alpha-\beta\vert$. We let $\left|0,0\right\rangle _{HG}$
denote the HG-mode fundamental Gaussian state. If we use the binomial
formula for Eq. $\left(\ref{eq:LG mode}\right)$, we can find a
more explicit form of LG-mode pointer states as a sum of HG modes:
\begin{equation}
\left|\mu,\nu\right\rangle _{LG}=\sum_{j=0}^{\alpha}\sum_{k=0}^{\beta}C_{\alpha,j;\beta,k}\left|\alpha+\beta-k-j,k+j\right\rangle _{HG}.\label{eq:LGGF}
\end{equation}
Here, we note that $C_{\alpha,j;\beta,k}$ is given by 
\begin{eqnarray}
C_{\alpha,j;\beta,k} & =\left(\frac{1}{\sqrt{2}}\right)^{\alpha+\beta}\frac{(-1)^{k}(i)^{k+j}}{\sqrt{\alpha!\beta!}}\times \nonumber \\
 & \sqrt{(\alpha+\beta-k-j)!(k+j)!}\left(\begin{array}{c}
\alpha\\
j
\end{array}\right)\left(\begin{array}{c}
\beta\\
k
\end{array}\right).
\label{eq:LGXF}
\end{eqnarray}
In the present paper, we take the initial state of the LG-mode pointer as
$\left|\varphi_{i}\right\rangle =\left|\mu,\nu\right\rangle _{LG}$.

The LG modes are a complete set of solutions to the paraxial wave
equation in cylindrical coordinates characterized by radial and
azimuthal indexes $p$ and $l$~\cite{Siegman(1986)}. Physically,
the LG modes have been created using various experimental setups such
as spatial light modulators \cite{Ando(2009)} and reflection from
a conical mirror \cite{Kobayashi(2012)}. Furthermore, the LG modes
have a zero-intensity point at the center called the optical
vortex. The relationship between the optical vortex and the weak value
has been investigated from different perspectives \cite{Shikano(2014),Denns(2012),Dnns(2012),Gotte(2012),Dennis,Magana(2013)}.
Thus, a general treatment of the post-selected von Neumann measurements
with LG-mode pointer states will provide an efficient method for further
exploration of weak-value applications in higher-order optical beams
and optical vortices. Next, we present an explicit treatment of
post-selected von Neumann measurements with LG-mode pointer states
for the system operator $\hat{A}$ that satisfies the properties $\hat{A}^{2}=\hat{I}$
and $\hat{A}^{2}=\hat{A}$.

\subsection{$\hat{A}^{2}=\hat{I}$ case}
By using the same process as that used in the HG-mode cases, after the
unitary evolution given in Eq. $\left(\ref{eq:UNA1}\right)$ and the
post-selection of the system to $\left|\psi_{f}\right\rangle $, we
can obtain the normalized final-pointer states as 
\begin{equation}
\left|\varphi_{f_{1}}\right\rangle =\frac{\lambda^{\prime}}{2}\left[D\left(\frac{s}{2}\right)+D\left(-\frac{s}{2}\right)+\langle A\rangle_{w}\left\{ D\left(\frac{s}{2}\right)-D\left(-\frac{s}{2}\right)\right\} \right]\left|\mu,\nu\right\rangle _{LG},\label{eq:LGA21}
\end{equation}
where the normalization coefficient is given by 
\begin{eqnarray}
\lambda^{\prime} & = & \left[1+\frac{1}{2}\left(1-\vert\langle A\rangle_{w}\vert^{2}\right) \times \right. \nonumber \\ & & \left.
\left(e^{-\frac{s^{2}}{2}}\sum_{j,j^{\prime}=0}^{\alpha}\sum_{k,k^{\prime}=0}^{\beta}C_{\alpha,j;\beta,k}C_{\alpha,j^{\prime};\beta,k^{\prime}}^{\ast}\delta_{k^{\prime}+j^{\prime},k+j}L_{\alpha+\beta-k-j}(s^{2})-1\right) \right]^{-\frac{1}{2}}. \label{eq:LGA21C}
\end{eqnarray}

By using Eq. $\left(\ref{eq:LGA21}\right)$ and the displaced Fock states,
i.e., Eq. $\left(\ref{eq:DFS}\right)$, we can obtain the expectation
value of the position operator $\hat{X}$ under the final pointer
states $\left|\varphi_{f_{1}}\right\rangle $ as 
\begin{equation}
\langle X\rangle_{f_{1}}^{LG}=g\vert\lambda^{\prime}\vert^{2}\Re\langle A\rangle_{w}.\label{eq:LGA21X}
\end{equation}
Similarly, the expectation value of the momentum operator $\hat{P}_{x}$
under the final pointer states $\left|\varphi_{f_{1}}\right\rangle $
is given by 
\begin{eqnarray}
2g\langle P_{x}\rangle_{f_{1}}^{LG} & = & \vert\lambda^{\prime}\vert^{2}s^{2}\Im\langle A\rangle_{w}e^{-\frac{s^{2}}{4}}\sum_{j,j^{\prime}=0}^{\alpha}\sum_{k,k^{\prime}=0}^{\beta}C_{\alpha,j;\beta,k}C_{\alpha,j^{\prime};\beta,k^{\prime}}^{\ast}\delta_{k^{\prime}+j^{\prime},k+j} \times\nonumber \\
 &  & \sum_{l=0}^{\infty}\frac{(\alpha+\beta-k-j)!(-\frac{s}{4}^{2})^{l-(\alpha+\beta-k-j)}}{l!} \times \nonumber \\ 
& & \ \ \ \ L_{\alpha+\beta-k-j}^{(l-(\alpha+\beta-k-j))}\left(\frac{s^{2}}{4}\right)L_{\alpha+\beta-k-j}^{(l+1-(\alpha+\beta-k-j))}\left(\frac{s^{2}}{4}\right).\label{eq:LGA21P}
\end{eqnarray}

From the definitions of the HG and LG modes in the Fock state representation,
i.e., Eqs. $\left(\ref{eq:HG mode}\right)$ and $\left(\ref{eq:LGGF}\right)$, respectively,
we can see that the LG modes are not factorable
into functions depending only on $x$ and $y$, in contrast to the HG modes. This feature of the
LG modes causes the coupling of the system observable $\hat{A}$ with
the $x$- and $y$-dimension of the pointer. Thus, the pointer also
shifts values in the $y$-direction. The pointer value is given by 
\begin{eqnarray}
\langle Y\rangle_{f_{1}}^{LG} & = & g\vert\lambda^{\prime}\vert^{2}\Im\langle A\rangle_{w}e^{-\frac{s^{2}}{2}}\sum_{j,j^{\prime}=0}^{\alpha}\sum_{k,k^{\prime}=0}^{\beta}\Re\{iC_{\alpha,j;\beta,k}C_{\alpha,j^{\prime};\beta,k^{\prime}}^{\ast}\} \times \nonumber \\ & & \ \ \ \ \delta_{k^{\prime}+j^{\prime},k+j-1}\sqrt{\frac{k+j}{\left(\alpha+\beta-k-j+1\right)}}L_{\alpha+\beta-k-j}^{\left(1\right)}\left(s^{2}\right)\nonumber \\
 &  & -g\vert\lambda^{\prime}\vert^{2}\Im\langle A\rangle_{w}e^{-\frac{s^{2}}{2}}\sum_{j,j^{\prime}=0}^{\alpha}\sum_{k,k^{\prime}=0}^{\beta}\Re\{iC_{\alpha,j;\beta,k}C_{\alpha,j^{\prime};\beta,k^{\prime}}^{\ast}\} \times \nonumber \\ & & \ \ \ \ \delta_{k^{\prime}+j^{\prime},k+j+1}\sqrt{\frac{k+j+1}{\left(\alpha+\beta-k-j\right)}}L_{\alpha+\beta-k-j-1}^{\left(1\right)}\left(s^{2}\right).\label{eq:LGA21Y}
\end{eqnarray}

These expectation values are the general forms of the desired values
in post-selected von Neumann measurements with LG pointer states for
the system operator $\hat{A}$ satisfying the property $\hat{A}^{2}=\hat{I}$.

\begin{figure}
\begin{center}
\centering{}\includegraphics[width=13cm]{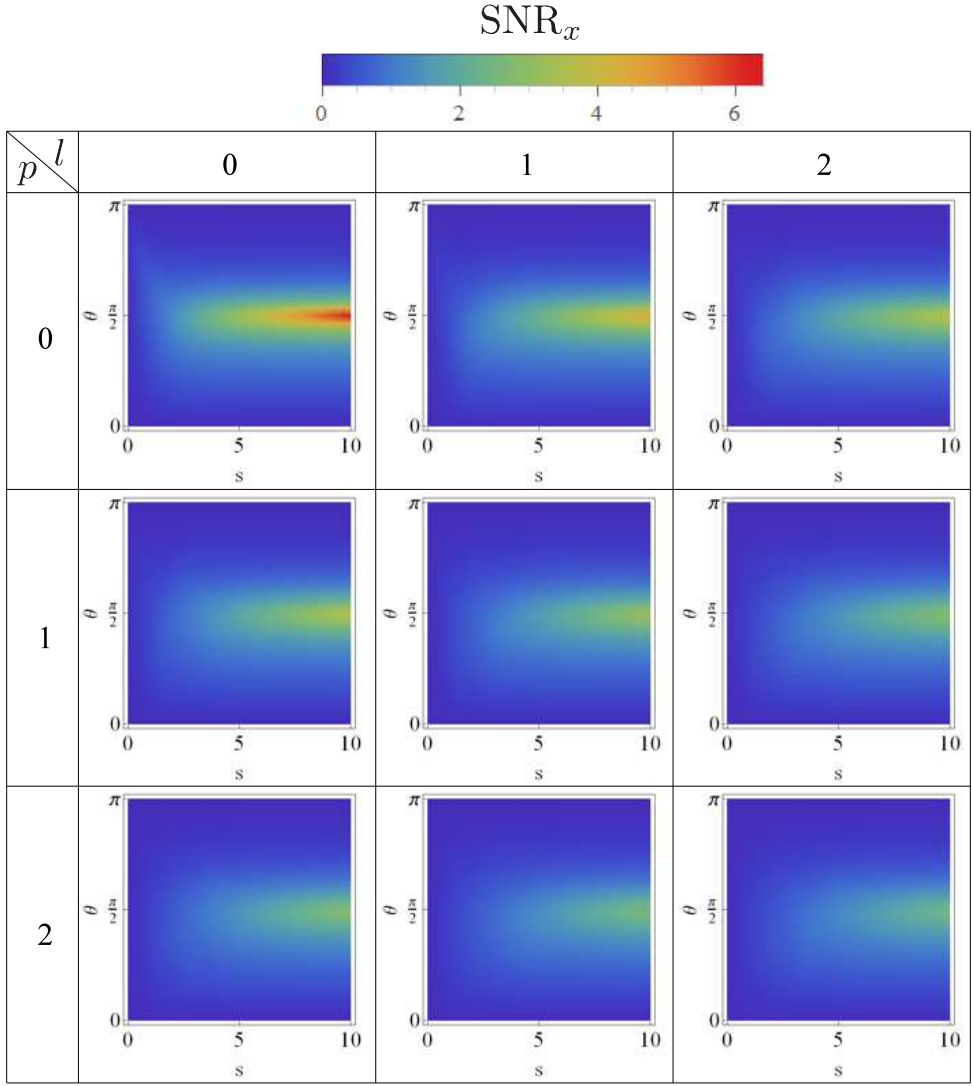} 
\end{center}
\protect\caption{(Color online)\label{fig:LGA21SNR} SNR in the $x$-direction
for LG-mode pointer states with the operator $\hat{A}$ satisfying
the property $\hat{A}^{2}=\hat{I}$ plotted with respect to the
measurement-strength parameter $s$ and pre-selection angle $\theta$
with $\phi=0$ fixed. These figures show the SNRs for
the lowest-order LG modes.}
\end{figure}

\begin{figure}
\begin{center}
\includegraphics[width=6cm]{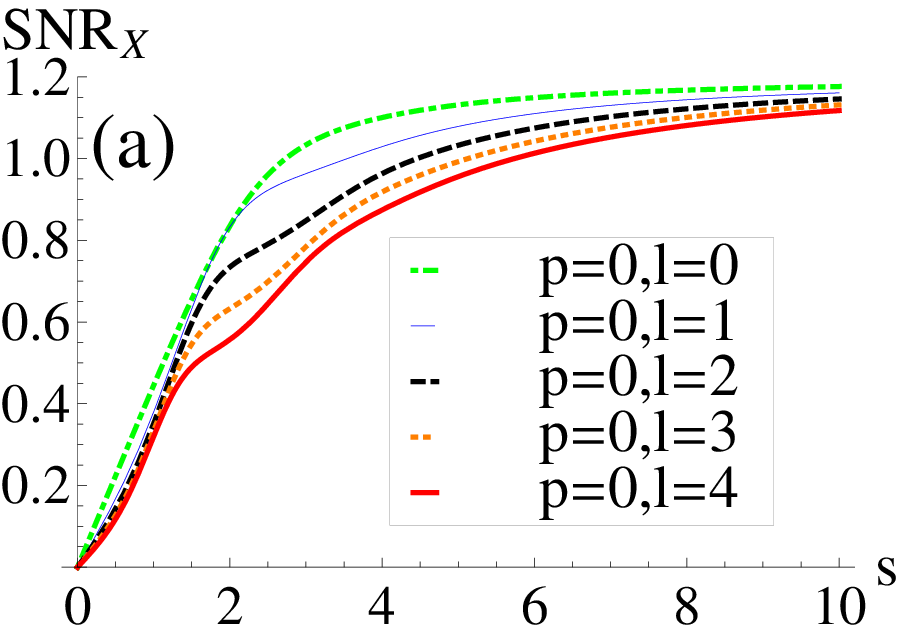} \ \includegraphics[width=6cm]{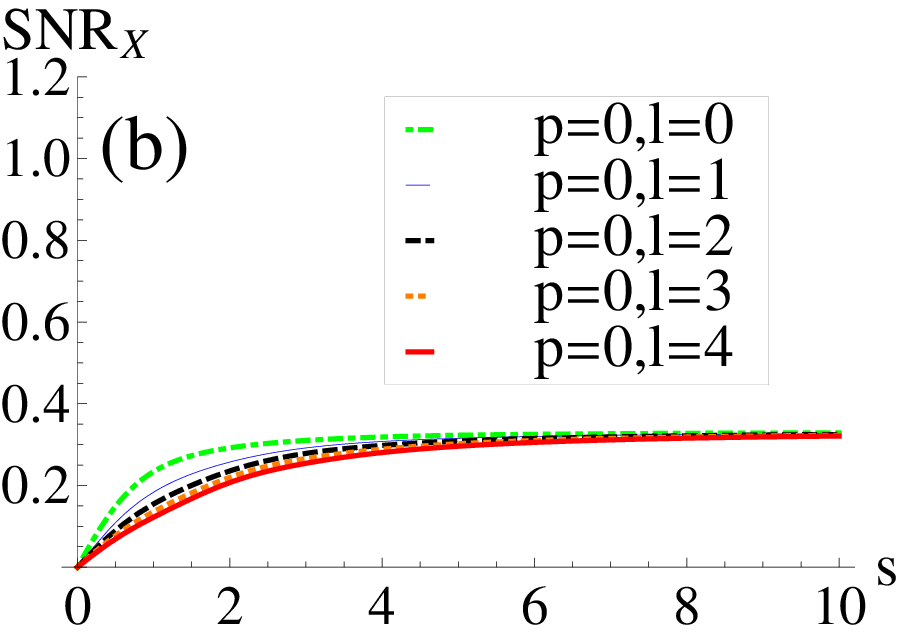}

\includegraphics[width=6cm]{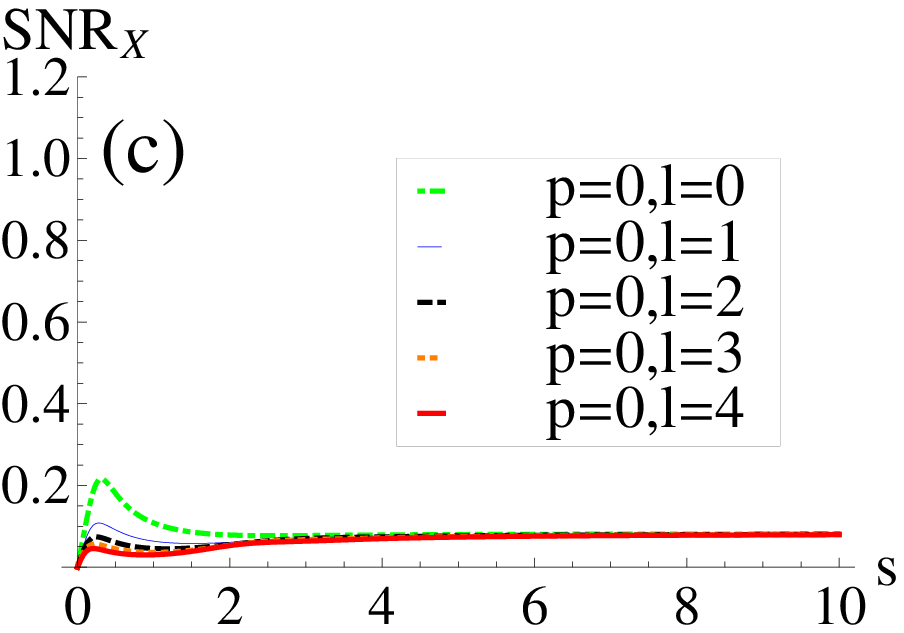} \ \includegraphics[width=6cm]{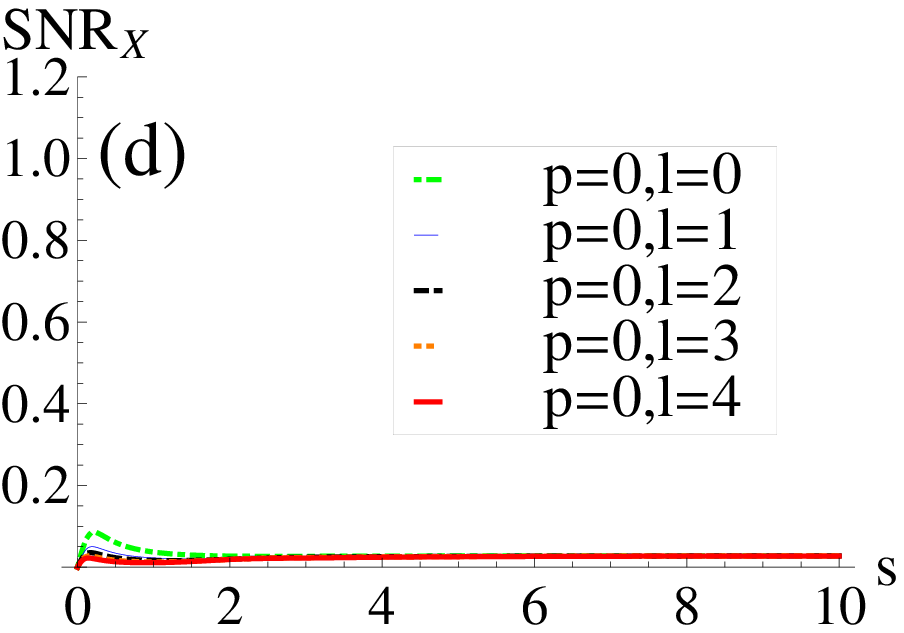}
\end{center}
\protect\caption{(Color online)\label{fig:SNRA21LG} SNR in the $x$-direction
for LG-mode pointer states with the operator $\hat{A}$ satisfying
the property $\hat{A}^{2}=\hat{I}$ plotted with respect to the
measurement-strength parameter $s$ for some specific weak values:
(a) $\langle A\rangle_{w}=0.5$, (b) $\langle A\rangle_{w}=0.5+i$,
(c) $\langle A\rangle_{w}=5$, and (d) $\langle A\rangle_{w}=5+5i$~\cite{paraex}.}
\end{figure}

For the weak value $\left(\ref{eq:wv}\right)$ with $\phi=0$ fixed,
the SNR is determined to be a function of the coupling parameter $s$
and the pre-selection angle $\theta$ for lower radial and azimuthal
indices $p$ and $l$, respectively, as shown in Fig. $\ref{fig:LGA21SNR}$. In the figure,
we show plots only for $p=0,1,2$ and the corresponding $l=0,1,2$
cases. Furthermore, by selecting specific weak values, we plot the
SNR as a function of the measurement-strength parameter $s$, as shown
in Fig. $\ref{fig:SNRA21LG}$. From Figs. $\ref{fig:LGA21SNR}$ and
$\ref{fig:SNRA21LG}$, we can see that the higher-order LG modes have
no advantages in improving the SNR over the case of the fundamental Gaussian
mode (corresponding to the $p=0,$ $l=0$ case). From Fig. $\ref{fig:SNRA21LG}$,
we can also see that the imaginary part of the weak value has no role
in improving the SNR in the $x$-direction.

\begin{figure}[t]
\begin{center}
\includegraphics[width=13cm]{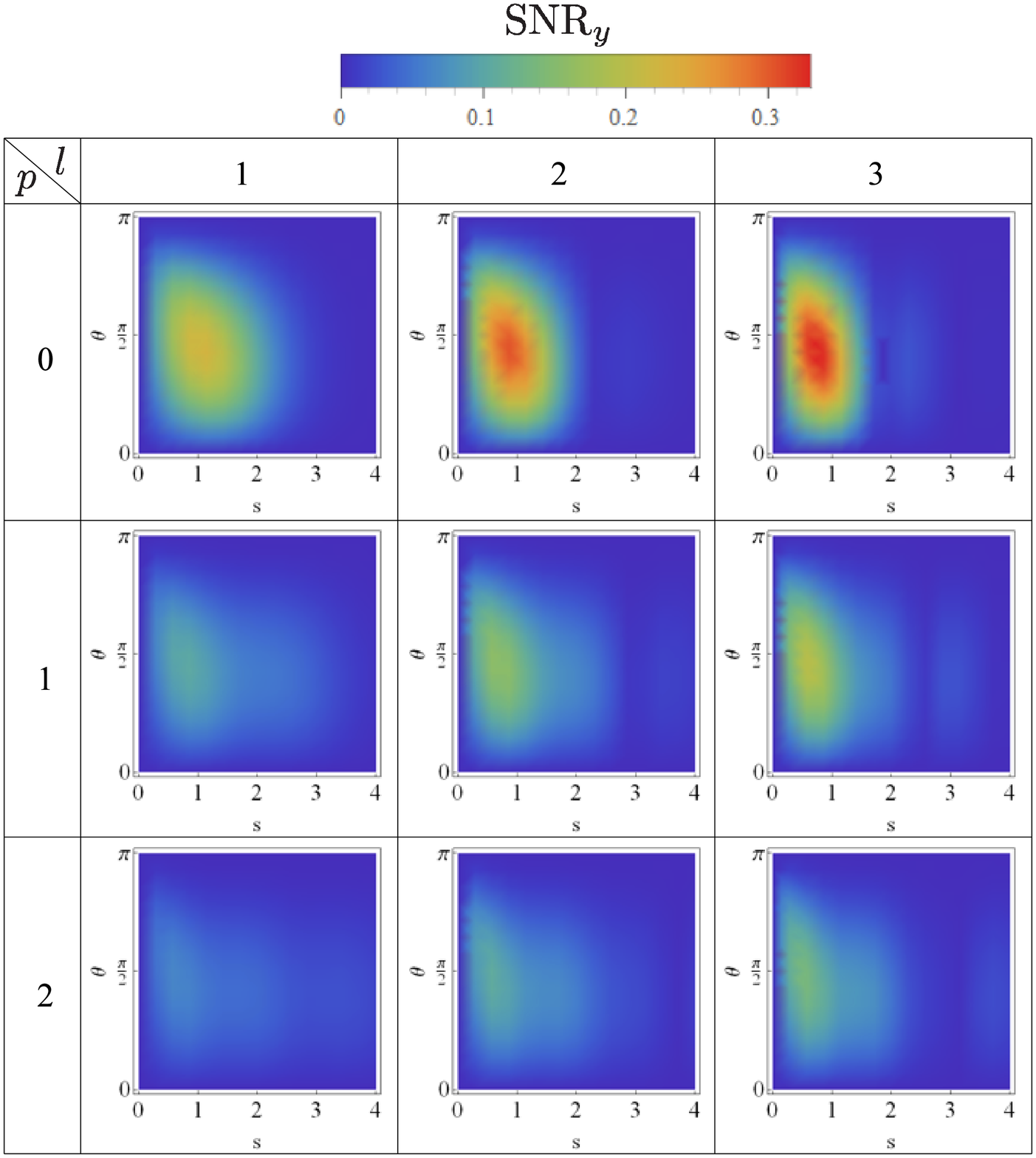} 
\end{center}
\protect\caption{(Color online) \label{fig:SNRy} SNR in the $y$-direction for
LG-mode pointer states with the operator $\hat{A}$ satisfying the
property $\hat{A}^{2}=\hat{I}$ plotted with respect to the measurement-strength parameter $s$. Here, we take $\phi=\frac{\pi}{2}$ in Eq. $\left(\ref{eq:wv}\right)$.}
\end{figure}

\begin{figure}
\begin{center}
\includegraphics[width=6cm]{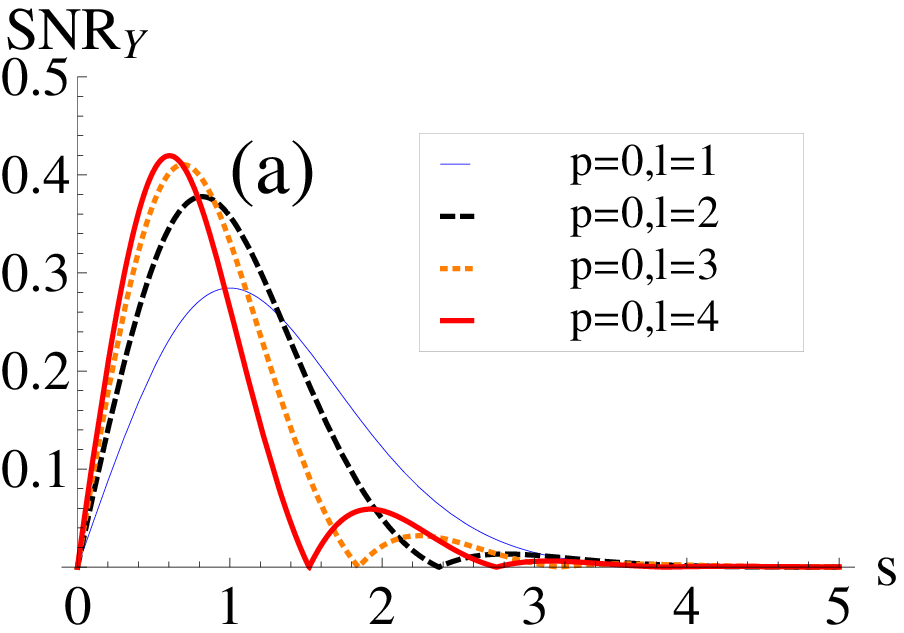} \ \includegraphics[width=6cm]{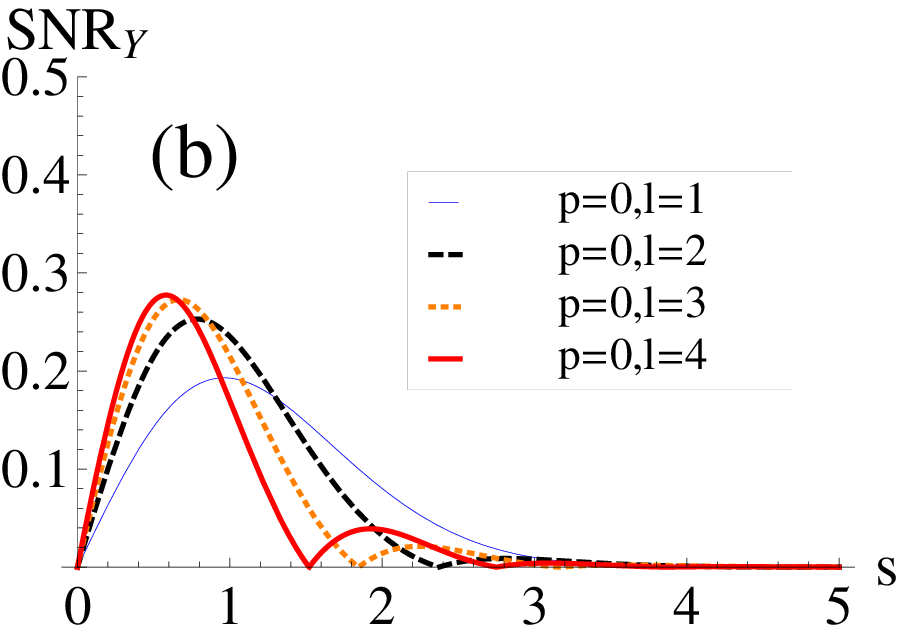}

\includegraphics[width=6cm]{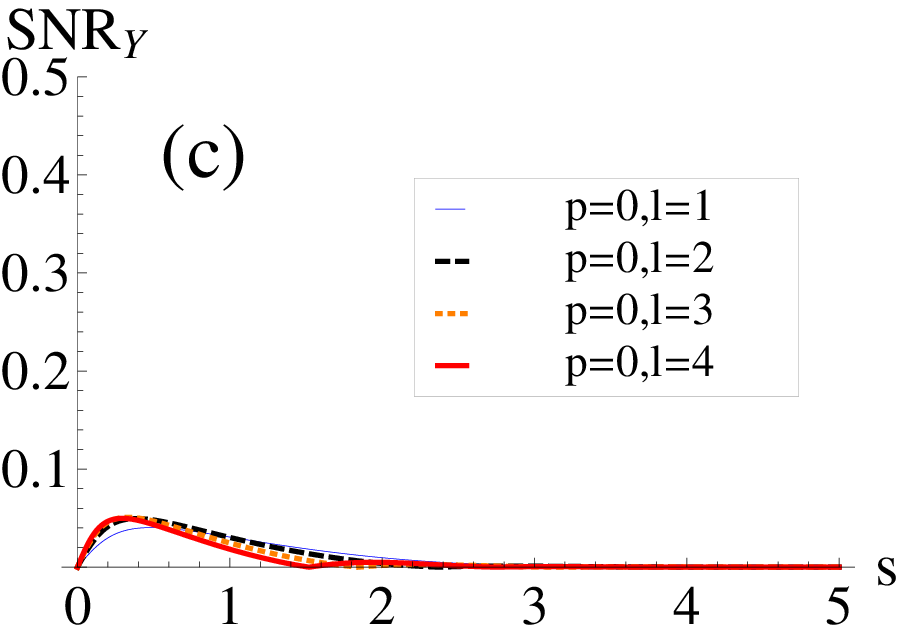} \ \includegraphics[width=6cm]{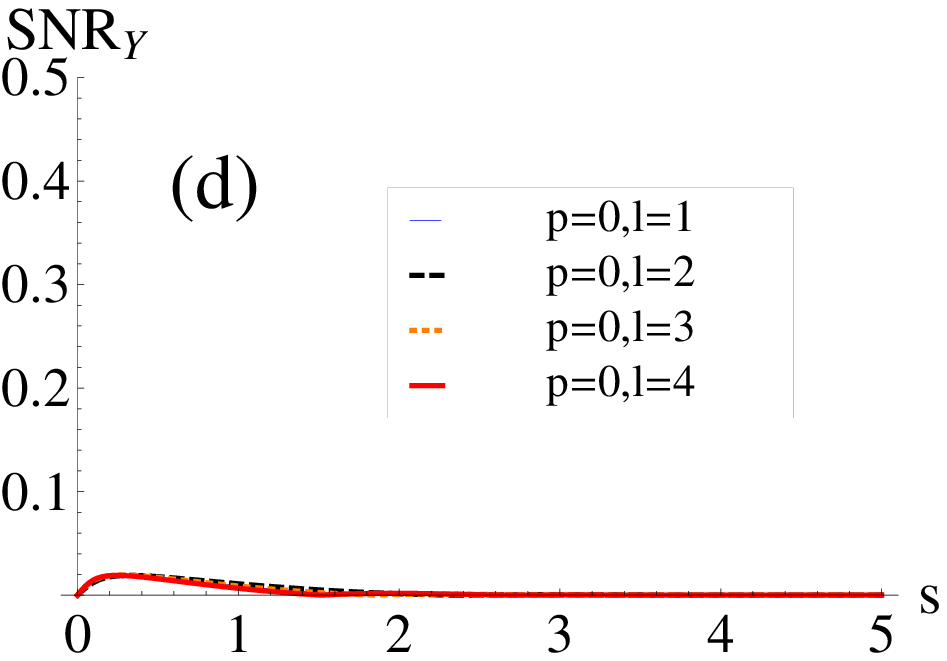}
\end{center}
\protect\caption{(Color online) \label{fig:LGA21y} SNR in the $y$-direction for
LG-mode pointer states with the operator $\hat{A}$ satisfying
the property $\hat{A}^{2}=\hat{I}$ for (a) $\langle A\rangle_{w}=i$,
(b) $\langle A\rangle_{w}=0.5+i$, (c) $\langle A\rangle_{w}=5i$,
and (d) $\langle A\rangle_{w}=5+5i$~\cite{paraex}.}
\end{figure}

The SNR for the $y$-direction shift is shown in Figs. \ref{fig:SNRy}
and \ref{fig:LGA21y}. In Fig. \ref{fig:SNRy}, the SNRs for
the lower-order cases of LG modes are shown, while in Fig.\ref{fig:LGA21y},
the SNR in the $y$-direction is plotted as a function of the measurement-strength parameter $s$ for some specific weak values with the radial
index fixed at $p=0$ and azimuthal index $l$ increasing.
We can observe that the SNR in the $y$-direction is related to the azimuthal
indices $l$, while the SNR decreases as the radial indices $p$ are
increased (see Fig. \ref{fig:SNRy}). Thus, when $l=0$, there is
no information about the $y$-direction. We should emphasize that
while the maximum of the $y$-direction shift is very small compared
to that of the $x$-direction shift, in the weak measurement regime, the
$y$-direction shift is sufficiently large compared to the $x$-direction
shift. In Fig. \ref{fig:LGA21y}, we can also observe that the real
part of the weak value has no role in improving the SNR in the $y$-direction.
Because there is no direct interaction between the pointer and the
measured system along the $y$-direction, the strong measurement regime
($s\gg1$) includes only the $x$-direction shift. In the weak measurement
regime, however, the pointer state can be shifted along not only the
$x$-direction but also the $y$-direction because the unfactorability
of the LG modes induces $y$-direction interference for $x$-direction
interaction. On the improvement of the SNR in the $y$-direction, it seems 
to be converged to the specific value on increasing the azimuthal
indices $l$.

\subsection{$\hat{A}^{2}=\hat{A}$ case}
Using a process similar to that in the previous section, we can determine
the normalized final state of the LG-mode pointer states as follows: 
\begin{equation}
\left|\varphi_{f_{2}}\right\rangle =\gamma^{\prime}\left[1-\langle A\rangle_{w}+\langle A\rangle_{w}D\left(\frac{g}{2\sigma}\right)\right]\left|\mu,\nu\right\rangle _{LG},\label{eq:LGA2AF}
\end{equation}
for the normalization coefficient 
\begin{eqnarray}
\gamma^{\prime} & = & \left[1+2\left(\Re\langle A\rangle_{w}-\vert\langle A\rangle_{w}\vert^{2}\right) \times \right. \nonumber \\ & & \left. \left(e^{-\frac{s^{2}}{8}}\sum_{j,j^{\prime}=0}^{\alpha}\sum_{k,k^{\prime}=0}^{\beta}C_{\alpha,j;\beta,k}C_{\alpha,j^{\prime};\beta,k^{\prime}}^{\ast}\delta_{k^{\prime}+j^{\prime},k+j}L_{\alpha+\beta-k-j}\left(\frac{s^{2}}{4}\right)-1\right)\right]^{-\frac{1}{2}}. \label{eq:LGA2AC}
\end{eqnarray}
The expectation values of the position operators $\hat{X}$, $\hat{Y}$,
and the momentum operator $\hat{P}_{x}$ under the final state $\left|\varphi_{f_{2}}\right\rangle $
are given by 
\begin{eqnarray}
\langle X\rangle_{f_{2}}^{LG} & = & \vert\gamma^{\prime}\vert^{2}g\left(\Re\langle A\rangle_{w}-\vert\langle A\rangle_{w}\vert^{2}\right)e^{-\frac{s^{2}}{8}} \times \nonumber \\ & & \sum_{j,j^{\prime}=0}^{\alpha}\sum_{k,k^{\prime}=0}^{\beta}C_{\alpha,j;\beta,k}C_{\alpha,j^{\prime};\beta,k^{\prime}}^{\ast}\delta_{k^{\prime}+j^{\prime},k+j}L_{\alpha+\beta-k-j}\left(\frac{s^{2}}{4}\right) \nonumber \\ & & +g\vert\gamma^{\prime}\vert^{2}\vert\langle A\rangle_{w}\vert^{2}, \label{eq:LGA2AX} \\
\langle Y\rangle_{f_{2}}^{LG} & = & -g\vert\gamma^{\prime}\vert^{2}\Im\langle A\rangle_{w}e^{-\frac{s^{2}}{8}}\sum_{j,j^{\prime}=0}^{\alpha}\sum_{k,k^{\prime}=0}^{\beta}\Re\left\{ iC_{\alpha,j;\beta,k}C_{\alpha,j^{\prime};\beta,k^{\prime}}^{\ast}\right\} \times \nonumber \\ & & \delta_{k^{\prime}+j^{\prime},k+j+1}\sqrt{\frac{k+j+1}{\alpha+\beta-k-j}}L_{\alpha+\beta-k-j-1}^{(1)}\left(\frac{s^{2}}{4}\right)\nonumber \\
 &  & +g\vert\gamma^{\prime}\vert^{2}\Im\langle A\rangle_{w}e^{-\frac{s^{2}}{8}}\sum_{j,j^{\prime}=0}^{\alpha}\sum_{k,k^{\prime}=0}^{\beta}\Re\left\{ iC_{\alpha,j;\beta,k}C_{\alpha,j^{\prime};\beta,k^{\prime}}^{\ast}\right\} \times \nonumber \\ & & \delta_{k^{\prime}+j^{\prime},k+j-1}\sqrt{\frac{k+j}{\alpha+\beta-k-j+1}}L_{\alpha+\beta-k-j}^{(1)}\left(\frac{s^{2}}{4}\right),\label{eq:LGA2AY}
\end{eqnarray}
and 
\begin{eqnarray}
2g\langle P_{x}\rangle_{f_{2}}^{LG} & = & \vert\gamma^{\prime}\vert^{2}s^{2}\Im\langle A\rangle_{w}e^{-\frac{s^{2}}{8}}
\sum_{j,j^{\prime}=0}^{\alpha}\sum_{k,k^{\prime}=0}^{\beta}C_{\alpha,j;\beta,k}C_{\alpha,j^{\prime};\beta,k^{\prime}}^{\ast} \times\nonumber \\ & & 
\delta_{k^{\prime}+j^{\prime},k+j}\left[L_{\alpha+\beta-k-j}^{(1)}\left(\frac{s^{2}}{4}\right)+L_{\alpha+\beta-k-j-1}^{(1)}\left(\frac{s^{2}}{4}\right)\right],\label{eq:LGA2AP}
\end{eqnarray}
respectively. 

\begin{figure}
\begin{center}
\includegraphics[width=6cm]{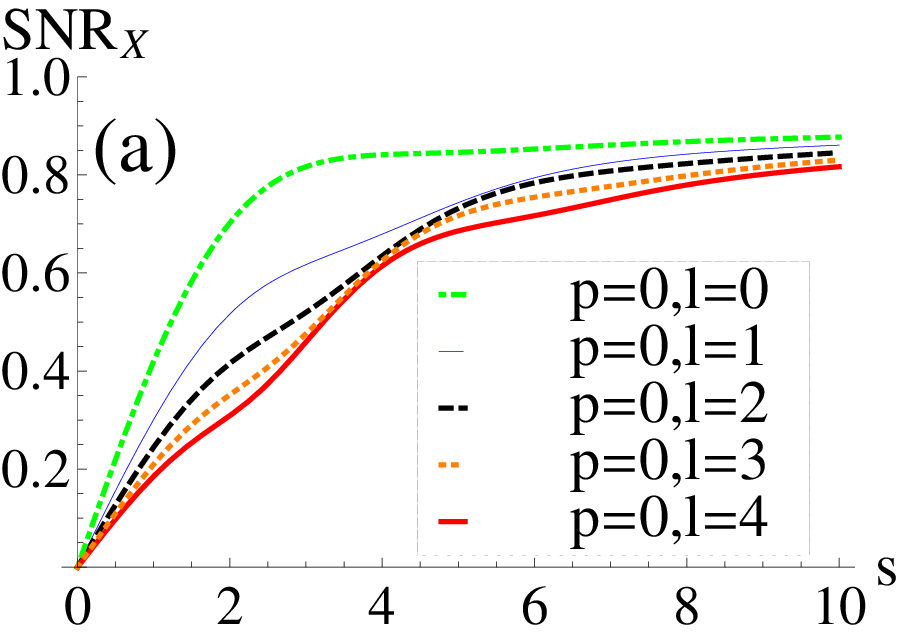} \ \includegraphics[width=6cm]{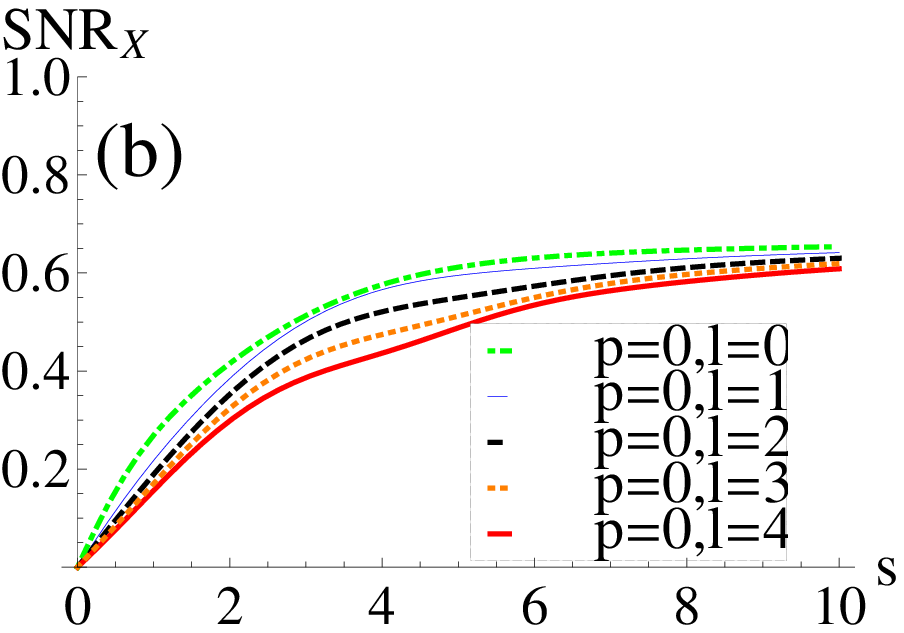}

\includegraphics[width=6cm]{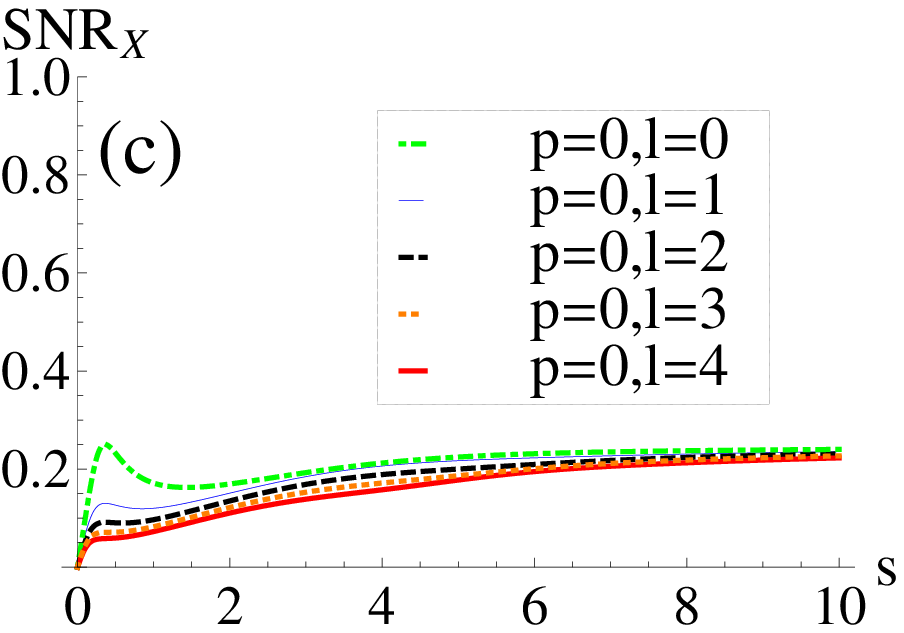} \ \includegraphics[width=6cm]{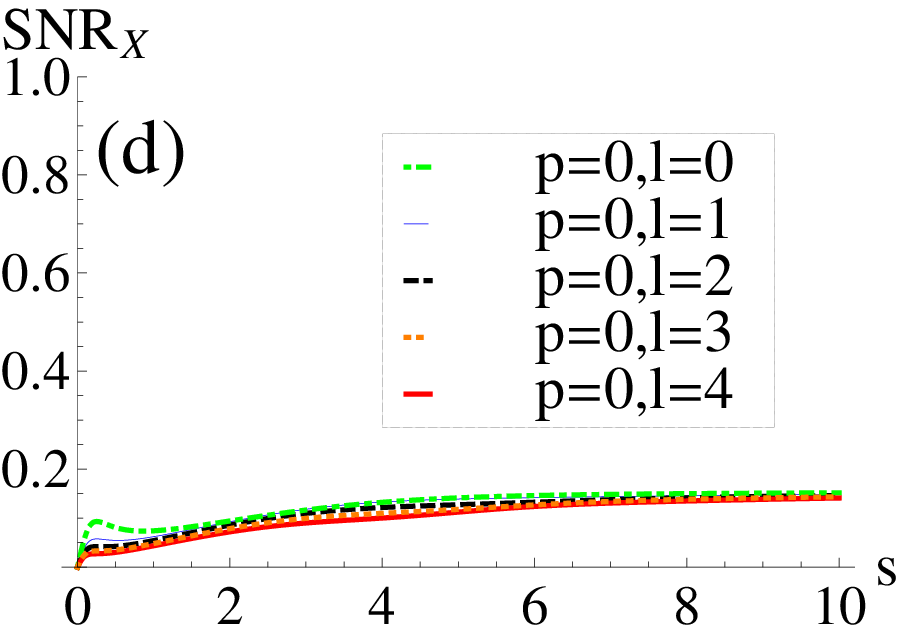}
\end{center}
\protect\caption{(Color online)\label{fig:SNRLGA2A} SNR in the $x$-direction
for LG-mode pointer states with the operator $\hat{A}$ satisfying
the property $\hat{A}^{2}=\hat{A}$ plotted with respect to the
measurement-strength parameter $s$ for specific weak values: (a)
$\langle A\rangle_{w}=0.5$, (b) $\langle A\rangle_{w}=0.5+i$, (c)
$\langle A\rangle_{w}=5$, and (d) $\langle A\rangle_{w}=5+5i$~\cite{paraex}.}
\end{figure}

\begin{figure}
\begin{center}
\includegraphics[width=4cm]{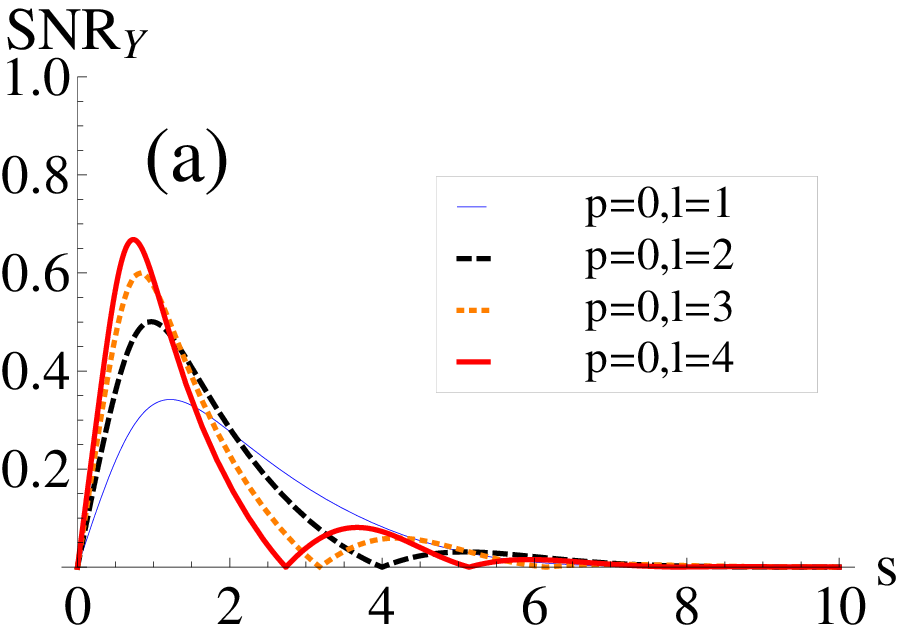} \ \includegraphics[width=4cm]{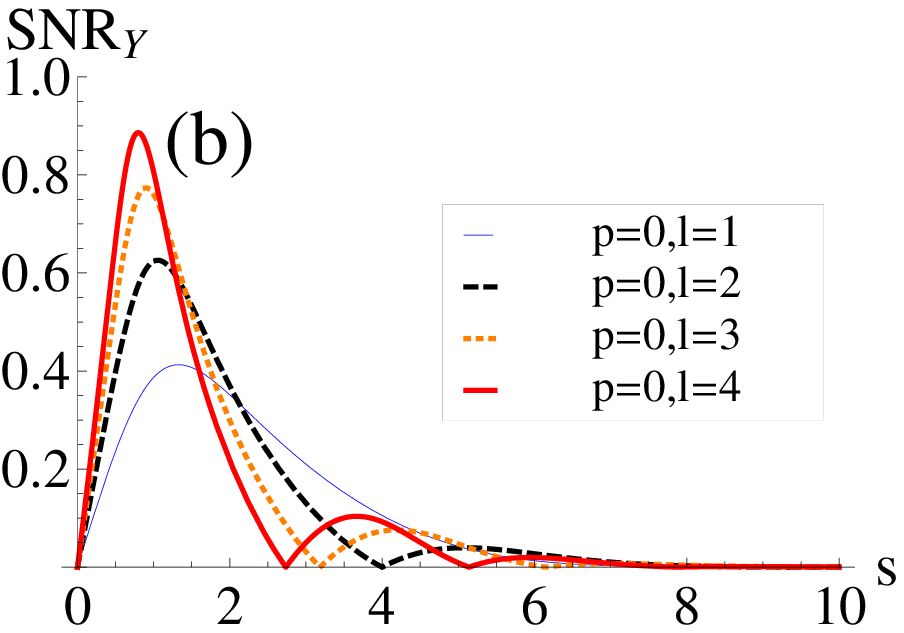} \ \includegraphics[width=4cm]{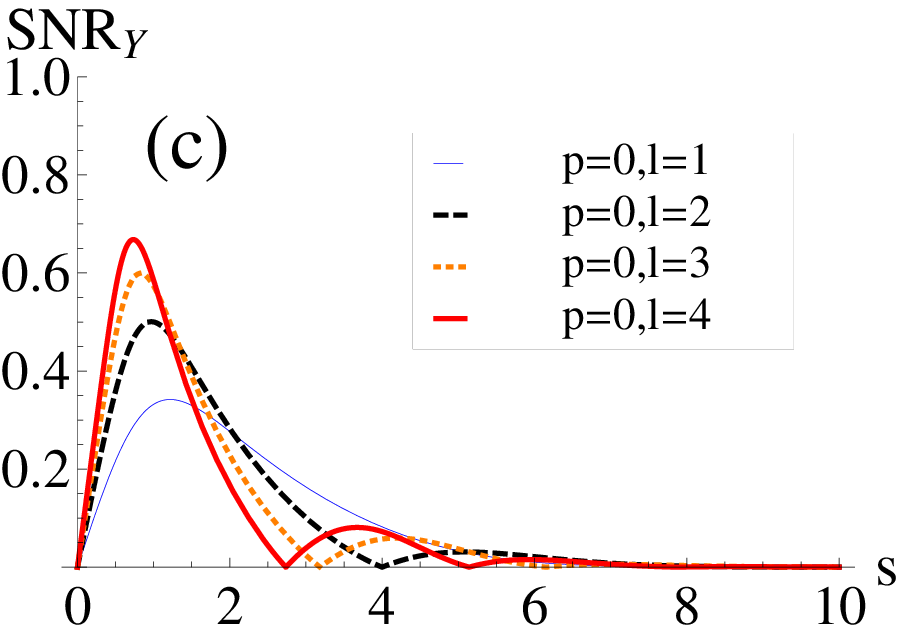}

\includegraphics[width=4cm]{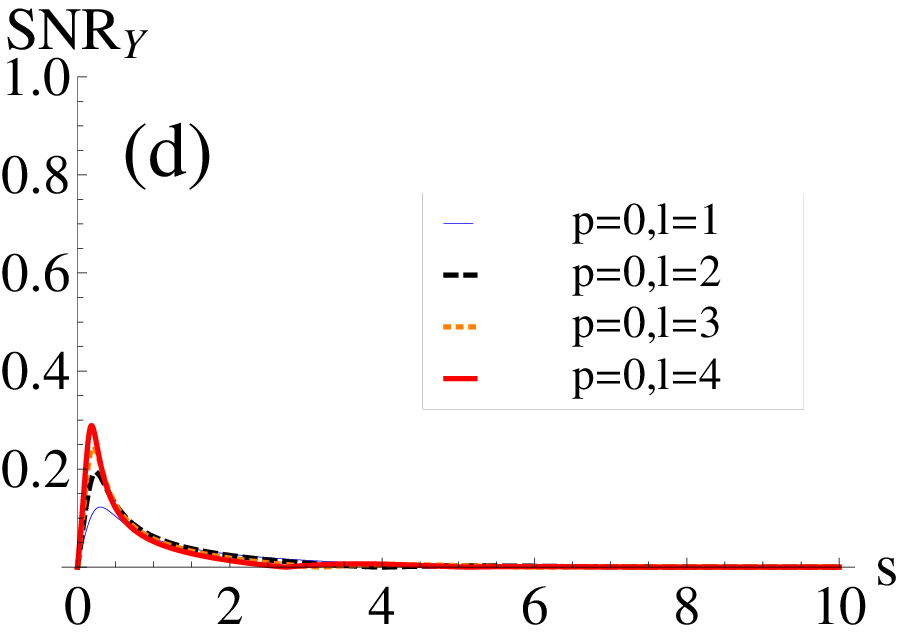} \ \includegraphics[width=4cm]{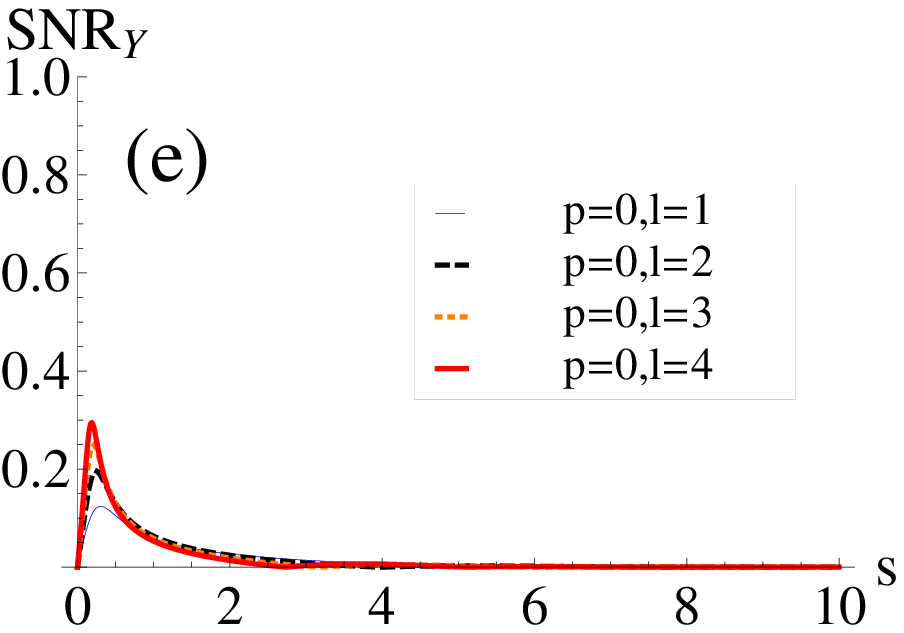} \ \includegraphics[width=4cm]{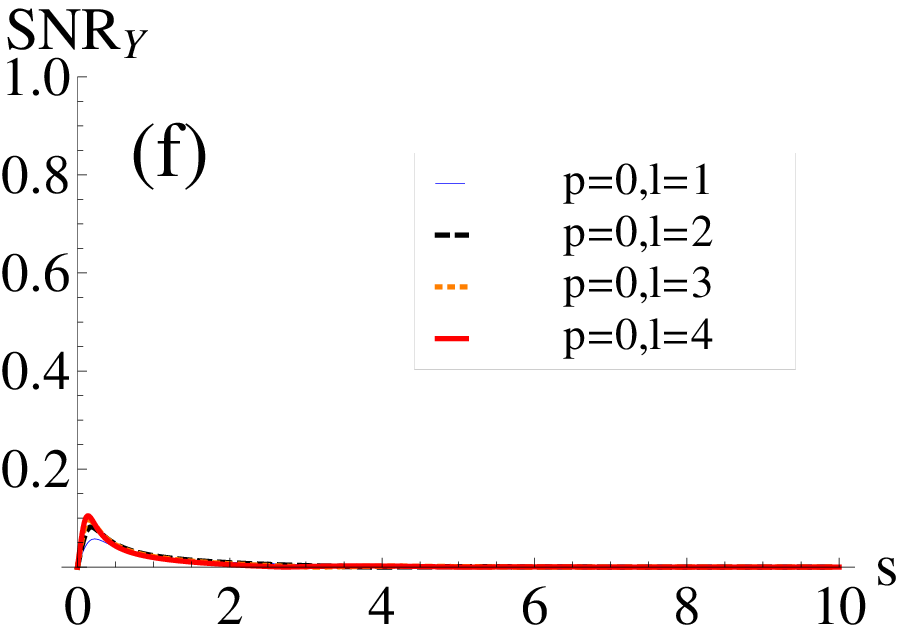}
\end{center}
\protect\caption{(Color online) \label{fig:SNRA2AY} SNR in the $y$-direction for
LG-mode pointer states with the operator $\hat{A}$ satisfying
the property $\hat{A}^{2}=\hat{A}$ plotted with respect to the
measurement-strength parameter $s$ for specific weak values: (a)
$\langle A\rangle_{w}=i$, (b) $\langle A\rangle_{w}=0.5+i$, (c)
$\langle A\rangle_{w}=1+i$, (d) $\langle A\rangle_{w}=5i$, (e) $\langle A\rangle_{w}=0.5+5i$,
and (f) $\langle A\rangle_{w}=5+5i$~\cite{paraex}.}
\end{figure}

For LG-mode pointer states with the system operator $\hat{A}$ satisfying
the property $\hat{A}^{2}=\hat{A}$, we verify the SNR values in the $x$-
and $y$-direction as functions of measurement-strength parameter $s$
with some specific weak values, and the numerical results are given
in Fig. $\ref{fig:SNRLGA2A}$ and Fig. \ref{fig:SNRA2AY}, respectively.
For the SNR in the $x$-direction, we reach the same conclusions as before:
the higher-order LG modes and imaginary parts of the weak value have
no advantages in improving the SNR in the $x$-direction (see Fig.
\ref{fig:SNRLGA2A}).

In Fig. \ref{fig:SNRA2AY},
we plot the SNR curves in the $y$-direction with the radial index fixed at
$p=0$ and azimuthal index $l$ changing. From Fig. \ref{fig:SNRA2AY},
we can observe that in the weak measurement regime ($s\ll1$), the SNR in the $y$-direction is improved
in comparison with the case $\hat{A}^{2}=\hat{I}$
shown in Fig. \ref{fig:LGA21y}. We numerically find that the maximum value
of the SNR occurs for $\langle A\rangle_{w}=0.5+i$, as shown in Fig.\ref{fig:SNRA2AY}(e). 
The maximum condition of this
SNR corresponds to the minimum condition for Eq. (\ref{eq:LGA2AC}).
Furthermore, from Fig. \ref{fig:SNRA2AY}, we also can see that when
the azimuthal index $l$ increases, the SNR in the $y$-direction
increases for a fixed radial index $p$. When the coupling between
the system ($x$-direction) and the pointer devices is sufficiently
strong, the SNR in the $y$-direction gradually vanishes. From Fig. \ref{fig:SNRA2AY},
we can further deduce that the real part of the weak value has no
role in improving the SNR in the $y$-direction. Note that these results 
investigate the importance of the imaginary part of the weak value such as Refs.~\cite{Knee(2014)-2,Justin}.
Also, there still is the open problem whether the unified information of the $x-$ and $y-$ directions is useful
as the optical implementation of the parameter estimation. 

\section{Some approximation cases}
\label{sec5} 
\subsection{$\hat{A}^{2}=\hat{I}$ case}
If we take the fundamental Gaussian beam as the initial pointer state
(this corresponds to taking $m=n=0$ and $\alpha=\beta=0$ in Eq.
$\left(\ref{eq:HG mode}\right)$ and Eq. $\left(\ref{eq:LG mode}\right)$,
respectively), the general expectation values for position operator
$\hat{X}$, i.e., Eqs. $\left(\ref{eq:HGGXA21}\right)$ and $\left(\ref{eq:LGA21X}\right)$,
and momentum operator $\hat{P}$, i.e., Eqs. $\left(\ref{eq:HGGPA21}\right)$
and $\left(\ref{eq:LGA21P}\right)$, are reduced to

\begin{equation}
\langle X\rangle_{f_{1},FG}=\frac{g\Re\langle A\rangle_{w}}{\mathit{\mathcal{Z}}}\label{eq:HGzerox}
\end{equation}
and 
\begin{equation}
\langle P_{x}\rangle_{f_{1},FG} = \frac{g\Im\langle A\rangle_{w}}{2\sigma^{2}\mathbb{\mathcal{Z}}}e^{-\frac{s^{2}}{2}},\label{eq:HGzerop}
\end{equation}
respectively, where 
\begin{equation}
\mathbb{\mathcal{Z}}=1+\frac{1}{2}\left(1-\vert\langle A\rangle_{w}\vert^{2}\right)\left(e^{-\frac{s^{2}}{2}}-1\right).\label{eq:zeroco}
\end{equation}
These results were also presented in Ref. \cite{Nakamura(2012)}.

Furthermore, under the weak measurement regime ($s\ll1$), if we only
consider evolution up to the first order, our general expectation
values reproduce the results given in Ref. \cite{deLima(2014)}.
In this case, the HG and LG pointer states are shifted along $x-$direction
with the same value, i.e., 
\begin{equation}
\langle X\rangle_{f_{1},first}=g\Re\langle A\rangle_{w}.\label{eq:HGfirA21}
\end{equation}
The expectation value in the $y$-direction, i.e., Eq.$\left(\ref{eq:LGA21Y}\right)$,
is reduced to 
\begin{equation}
\langle Y\rangle_{f_{1},first}^{LG}=-lg\Im\langle A\rangle_{w}.\label{eq:LGfirA21Y}
\end{equation}
The expectation value of the momentum operator for the HG-mode pointer
states, i.e., Eq. $\left(\ref{eq:HGGPA21}\right)$, is reduced to
\begin{equation}
\langle P_{x}\rangle_{f_{1},first}^{HG}=\frac{g\Im\langle A\rangle_{w}}{2\sigma^{2}}\left(2n+1\right),\label{eq:HGfirA21P}
\end{equation}
while that for the LG-mode pointer states, i.e., Eq. $\left(\ref{eq:LGA21P}\right)$,
is reduced to 
\begin{equation}
\langle P_{x}\rangle_{f_{1},first}^{LG}=\frac{g\Im\langle A\rangle_{w}}{2\sigma^{2}}\left(2p+\vert l\vert+1\right).\label{eq:LGfirA21P}
\end{equation}
The validity conditions for Eqs. $\left(\ref{eq:HGfirA21}\right)$--$\left(\ref{eq:LGfirA21P}\right)$
are 
\begin{equation}
\frac{g\sqrt{2n+1}}{2\sigma}\max\left(1,\vert\langle A\rangle_{w}\vert\right)\ll1,\label{eq:CoFA21HG}
\end{equation}
for the HG-mode pointer states, and 
\begin{equation}
\frac{g\sqrt{2p+\vert l\vert+1}}{2\sigma}\max\left(1,\vert\langle A\rangle_{w}\vert\right)\ll1,\label{eq:CoFA21LG}
\end{equation}
for the LG-mode pointer states.

The SNRs are directly related to measurement-strength
parameter $s$. Thus, in the strong measurement regime, if we take
the limit $s\rightarrow\infty$, we notice that $SNR_{X}$ becomes
a function of the weak value 
\begin{equation}
(SNR_{X})_{s\rightarrow\infty}=\frac{2\sqrt{P_{s}}\vert\Re\langle A\rangle_{w}\vert}{\sqrt{1+2\vert\langle A\rangle_{w}\vert^{2}+\vert\langle A\rangle_{w}\vert^{4}-4\Re^{2}\langle A\rangle_{w}}}.\label{eq:LmA21}
\end{equation}
We can observe this limiting trend from Figs. $\ref{fig:SNRA21HG2D}$
and $\ref{fig:SNRA21LG}$.

\subsection{$\hat{A}^{2}=\hat{A}$ case}
If we take the fundamental Gaussian beam as the initial pointer states,
the general expectation values for the position operator $\hat{X}$,
i.e., Eqs. $\left(\ref{eq:HGA2AX}\right)$ and $\left(\ref{eq:LGA2AX}\right)$,
and momentum operator $\hat{P}_{x}$, i.e., Eqs. $\left(\ref{eq:HGA2AP}\right)$
and $\left(\ref{eq:LGA2AP}\right)$, are reduced to 
\begin{equation}
\langle X\rangle_{f_{2},FG}=g\frac{\vert\langle A\rangle_{w}\vert^{2}+\left(\Re\langle A\rangle_{w}-\vert\langle A\rangle_{w}\vert^{2}\right)e^{-\frac{s^{2}}{8}}}{\mathbb{\mathcal{N}}},\label{eq:A2AFirX}
\end{equation}
and 
\begin{equation}
\langle P_{x}\rangle_{f_{2},FG}=\frac{g\Im\langle A\rangle_{w}}{2\sigma^{2}\mathbb{\mathcal{N}}}e^{-\frac{s^{2}}{8}},\label{eq:A2AFirP}
\end{equation}
respectively, where 
\begin{equation}
\mathbb{\mathcal{N}}=1+2\left(\Re\langle A\rangle_{w}-\vert\langle A\rangle_{w}\vert^{2}\right)\left(e^{-\frac{s{}^{2}}{8}}-1\right).\label{eq:COA2A}
\end{equation}
Furthermore, under the weak measurement regime ($s\ll1$), if we only
consider evolution up to the first order, our general expectation
values are reduced to the following form: 
\begin{equation}
\langle X\rangle_{f_{2},first}=g\Re\langle A\rangle_{w}.\label{eq:FA2ALHGX}
\end{equation}
In the case of the position operator $\hat{X}$, the HG mode and LG mode
have the same value. The expectation value in the $y$-direction,
Eq. $\left(\ref{eq:LGA2AY}\right)$, is reduced to 
\begin{equation}
\langle Y\rangle_{f_{2},first}^{LG}=-lg\Im\langle A\rangle_{w}.\label{eq:FA2AY}
\end{equation}
For the momentum operator $\hat{P}_{x}$, the expectation values for
the HG-mode pointer states, Eq. $\left(\ref{eq:HGA2AP}\right)$, and
for the LG-mode pointer states, Eq. $\left(\ref{eq:LGA2AP}\right)$,
are reduced to 
\begin{equation}
\langle P_{x}\rangle_{f_{2},first}^{HG}=\frac{g\Im\langle A\rangle_{w}}{2\sigma^{2}}\left(2n+1\right)\label{eq:FA2APHG}
\end{equation}
and 
\begin{equation}
\langle P_{x}\rangle_{f_{2},first}^{LG}=\frac{g\Im\langle A\rangle_{w}}{2\sigma^{2}}\left(2p+\vert l\vert+1\right),\label{eq:FA2APLG}
\end{equation}
respectively. The validity conditions for Eqs. $\left(\ref{eq:FA2ALHGX}\right)$--$\left(\ref{eq:FA2APLG}\right)$ are 
\begin{equation}
\frac{g\sqrt{2n+1}}{2\sigma}\max\left(1,\vert\langle A\rangle_{w}\vert,\sqrt{\vert\Re\langle A\rangle_{w}\vert}\right)\ll1,\label{eq:ConA2AHG}
\end{equation}
for the HG-mode pointer states and 
\begin{equation}
\frac{g\sqrt{2p+\vert l\vert+1}}{2\sigma}\max\left(1,\vert\langle A\rangle_{w}\vert,\sqrt{\vert\Re\langle A\rangle_{w}\vert}\right)\ll1,\label{eq:ConFA2ALG}
\end{equation}
for the LG-mode pointer states.

For the SNR in the strong-measurement regime ($s\gg1$), if we consider
the limiting case of $s\rightarrow\infty$, we note that $SNR_{X}$
becomes a function of the weak value 
\begin{equation}
\left(SNR_{X}\right)_{s\rightarrow\infty}=\frac{\sqrt{P_{s}}\vert\langle A\rangle_{w}\vert}{\sqrt{1+\vert\langle A\rangle_{w}\vert^{2}-2\Re\langle A\rangle_{w}}}.\label{eq:LmA2A}
\end{equation}
We can observe this limiting trend from Figs. $\ref{fig:SNR A2AHG}$
and $\ref{fig:SNRLGA2A}$.

We emphasize here that the limiting values given in Eqs. $\left(\ref{eq:LmA21}\right)$
and $\left(\ref{eq:LmA2A}\right)$ are valid in the $x$-direction
SNR for the HG- and LG-mode pointer states in corresponding lower-order
modes. It is assumed that the probe wavefunction does not spread out
during the interaction. Thus, for the case of the fundamental Gaussian
pointer, the expectation values of the position operator 
$\left(\ref{eq:HGzerox}, \ref{eq:A2AFirX} \right)$ 
and its conjugate momentum operator $\left(\ref{eq:HGzerop}, \ref{eq:A2AFirP} \right)$ 
are the same as those in Ref. \cite{Jozsa} under the weak-measurement condition.

\section{Conclusion and remarks}
\label{sec6} In summary, we studied the post-selected von Neumann measurement
with HG- and LG-mode pointer states for the system operator $\hat{A}$
satisfying $\hat{A}^{2}=\hat{I}$ and $\hat{A}^{2}=\hat{A}$. Our general
expectation formulas are valid in not only the weak-measurement regime
but also the strong-measurement regime. If we only consider
evaluation up to the first order, our general results reproduce all
results given in Ref. \cite{deLima(2014)}. Moreover, if we let the
initial pointer state be a fundamental Gaussian state, our general
results reflect the full evaluation values given in Ref. \cite{Nakamura(2012)}.

To clarify the practical advantages of high-order Gaussian beams,
we verified the SNR and found that the higher-order HG and LG
modes have no advantages for improving the SNR over that for
the case of the fundamental Gaussian mode. Moreover, we found that the imaginary
part of the weak values has no role in improving the SNR in the $x$-direction
in the cases of HG- and LG-mode pointer states. For the SNR in the $y$-direction
in the LG-mode case, we also found that the SNR is related to the
azimuthal index $l$ and that the real part of the weak value has no
role in improving the SNR in the $y$-direction. However, in the case of $\hat{A}^{2}=\hat{I}$,
the SNR in the $y$-direction has an upper bound even for increasing
azimuthal indices $l$. In the case of $\hat{A}^{2}=\hat{A}$, we observed
an improvement in SNR in the $y$-direction in the weak-measurement regime
because the SNR increases with increasing azimuthal index $l$. This fact 
may be helpful on the parameter estimation context as the optical implementation 
of the weak-value amplification.
However, we found that the SNR in the $y$-direction gradually vanishes
when the coupling strength between the system ($x$-direction) and
pointer devices is increased. It is noted that our choice of the pre- and post-selection 
may be not optimal to maximize the SNR. The SNR in the $y$-direction also disappeared in the weak-measurement
regime when the post-selected state is identical to the pre-selected one 
such that $\langle A\rangle_{w}=\langle\psi_{i}\vert A\vert\psi_{i}\rangle$.

These methods can provide a new technique for calculating the expectation
values of the generation functions of the momentum and position operators.
Thus, our results are useful for investigating applications of the
weak-measurement theory in quantum dynamics and quantum correlations
with higher-order optical beams. Also, these provide the role of the 
imaginary part of the weak value to lead to the complementarity relationship 
and the estimation problems in the Fourier domain for the LG higher order case.

We expect that our general treatment of the weak values will be helpful
for understanding the connection between weak- and strong-measurement
regimes and may be used to propose new experimental setups with higher-order
Gaussian beams to investigate further the applications of weak measurement
in optical systems such as the optical vortex. In this work, we only
consider the pure higher-order HG and LG modes as initial pointer
states and investigate the corresponding SNRs. However, the entanglement
of the initial pointer states~\cite{Pang(2014)} and the non-classical initial 
pointer states~\cite{Turek(2015)} are useful for the weak-value amplification. 
Thus, our setup may provide another scheme for
improving the SNR if we consider the initial state of the pointer
as a coherent-superposition state of higher-order Gaussian beams. 

\section*{Acknowledgments}
Y.T. would like to thank Taximaiti Yusufu for useful suggestions and
discussions. Y.S. thanks Shinji Yoshimura for discussions. This work
was supported by a Grant for Basic Science Research Projects from
The Sumitomo Foundation, a grant from Matsuo Foundation, IMS Joint
Study Program, NINS youth collaborative project, 
the Center for the Promotion of Integrated Sciences
(CPIS) of Sokendai, ICRR Joint Research from The University of Tokyo, 
and JSPS KAKENHI Grant Numbers 24654133, 25790068,
and 25287101. Y.T. acknowledges financial support from the IMS Internship
project.


\begin{thebibliography}{10}
\bibitem{paraex} In Eq. $\left(\ref{eq:wv}\right)$, $\langle A\rangle_{w} = 0.5$ is about $\phi = 0, \theta = 0.927$, $\langle A\rangle_{w} = i $ is $\phi = \pi/2, \theta = \pi /2$, $\langle A\rangle_{w} = 0.5 + i $ is about $\phi = 1.11, \theta = 1.68$, $\langle A\rangle_{w} = 1 + i $ is about $\phi = 0.785, \theta = 1.91$, $\langle A\rangle_{w} = 5$ is about $\phi = 0, \theta = 2.74$, $\langle A\rangle_{w} = 5i$ is about $\phi = \pi/2, \theta = 2.74$, and $\langle A\rangle_{w} = 5 + 5 i$ is about $\phi = 0.785, \theta = 2.86$.
\end{thebibliography}
\end{document}